\iffalse
\documentclass[aps,prl,twocolumn,
			   groupedaddress,superscriptaddress,
			   amsfonts,amssymb,amsmath,
			   citeautoscript,
			   a4paper]{revtex4-2}
\else
\documentclass[aps,pra,reprint,
			   groupedaddress,superscriptaddress,notitlepage,
			   amsfonts,amssymb,amsmath,
			   citeautoscript,
			   a4paper]{revtex4-2}
\fi

\usepackage[pdftex]{hyperref}
\hypersetup{colorlinks,
			linkcolor={blue!75!black!80!yellow},
			citecolor={blue!75!black!80!yellow},
			urlcolor={blue!75!black!80!yellow},
			pdfstartview=FitH}
\usepackage{graphicx}
\usepackage{mathrsfs}
\usepackage{amsthm}
\usepackage{xspace}
\usepackage{braket}
\usepackage{xr}
\usepackage{gensymb} 
\usepackage{xcolor,soul}
\usepackage{stmaryrd} 
\usepackage[UKenglish]{babel}
\usepackage{placeins} 
\usepackage{amsmath,amssymb}
\usepackage{siunitx}
\DeclareSIUnit[number-unit-product=]\percent{\char`\%} 

\usepackage{txfonts}  
\usepackage{txfontsb} 

\makeatletter
\newcommand*{\addFileDependency}[1]{
  \typeout{(#1)}
  \@addtofilelist{#1}
  \IfFileExists{#1}{}{\typeout{No file #1.}}
}
\makeatother

\makeatletter
\renewcommand\@make@capt@title[2]{%
	\@ifx@empty\float@link{\@firstofone}{\expandafter\href\expandafter{\float@link}}%
	\sffamily{\textbf{#1}}\@caption@fignum@sep#2
}%

\makeatother

\newcommand{\iu}{\mathrm{i}}
\newcommand{\e}{\mathrm{e}}

\newcommand{\appropto}{\mathrel{\vcenter{
			\offinterlineskip\halign{\hfil$##$\cr
				\propto\cr\noalign{\kern2pt}\sim\cr\noalign{\kern-2pt}}}}}

\newcommand{\sigmax}{\sigma_x}
\newcommand{\sigmay}{\sigma_y}
\newcommand{\sigmaz}{\sigma_z}

\usepackage{accents}

\usepackage{textcomp} 
\usepackage{xifthen}
\usepackage{etoolbox}
\newboolean{togglecomments}
\newboolean{togglechanges} 

\setboolean{togglecomments}{false}
\setboolean{togglechanges}{true}

\newcommand{\comment}[2]{%
    \ifbool{togglecomments}%
    {\textcolor{blue!70!black}{\small\textsf{%
    \textsuperscript{\textsc{\textsf{\MakeLowercase{#1}}}}%
    [#2]}}} 
    {}}     
\newcommand{\swap}[2]{\ifbool{togglechanges}
    {#2}  
    {\textcolor{red!70!black}{[#1]}\textrightarrow{}\textcolor{green!50!black}{[#2]}}}
\newcommand{\remove}[1]{\ifbool{togglechanges}
    {}    
    {\textcolor{red!70!black}{#1}}}
\newcommand{\inset}[1]{\ifbool{togglechanges}
    {#1}  
    {\textcolor{green!50!black}{#1}}}
\newcommand{\optional}[1]{\ifbool{togglechanges}
    {}    
    {\textcolor{yellow!50!orange!80!gray}{#1}}}

\newcommand{\citeremind}[1]{%
    [\textcolor{blue!75!black!80!yellow}{
        $\blacksquare$%
	    \ifthenelse{\isempty{#1}}
	        {}
	        {\textsuperscript{\tiny\textsf{#1}}}%
	}]\xspace}

\newcommand{\hkuaffil}{\footnotesize Department of Physics and HK Institute of Quantum Science and Technology, The University of Hong Kong, Pokfulam, Hong Kong, China}
\newcommand{\mitaffil}{\footnotesize Research Laboratory of Electronics and Department of Physics, Massachusetts Institute of Technology, Cambridge, Massachusetts 02139, USA}

\begin{document}

\title{Topological quantum walk in synthetic non-Abelian gauge fields}


\author{Zehai Pang}
\affiliation{\hkuaffil}
\author{Omar Abdelghani}
\affiliation{\mitaffil}
\author{Marin~Solja\v{c}i\'{c}}
\affiliation{\mitaffil}
\author{Yi~Yang}
\email{yiyg@hku.hk}
\affiliation{\hkuaffil}

\begin{abstract}
We theoretically introduce synthetic non-Abelian gauge fields for topological quantum walks. The photonic mesh lattice configuration is generalized with polarization multiplexing to achieve a four-dimensional Hilbert space, based on which we provide photonic building blocks for realizing various quantum walks in non-Abelian gauge fields.
It is found that SU(2) gauge fields can lead to Peierls substitution in both momenta and quasienergy. 
In one and two dimensions, we describe detailed photonic setups to realize topological quantum walk protocols whose Floquet winding numbers and Rudner--Lindner--Berg--Levin invariants can be effectively controlled by the gauge fields.
Finally, we show how non-Abelian gauge fields facilitate convenient simulation of entanglement in conjunction with polarization-dependent and spatial-mode-dependent coin operations. 
Our results shed light on the study of synthetic non-Abelian gauge fields in photonic Floquet systems. 
\end{abstract}

\maketitle

Photonics has recently become an emerging platform for exploring non-Abelian physics~\cite{yang2024non,chen2019non,yang2019synthesis,yang2020observation,guo2021experimental,jiang2021experimental,brosco2021non,chen2022classical,sun2022non,you2022observation,noh2020braiding,zhang2022non,xu2016simulating,yang20202d,yang2022non,gianfrate2020direct,whittaker2021optical,iadecola2016non,polimeno2021experimental,cheng2023artificial,bouhon2020non,jiang2024observation}. Light propagating in complex media offers versatile degrees of freedom such as polarization, angular momenta, and frequency modes that enable the realization of many non-Abelian phenomena, such as non-Abelian band topology~\cite{yang2020observation,guo2021experimental}, non-Abelian gauge fields~\cite{cheng2023artificial,yang2020non,yang2019synthesis}, and non-Abelian pumping~\cite{sun2022non,zhang2022non}.
In particular, non-Abelian gauge fields have been created on anisotropic, polaritonic, and waveguide systems, leading to synthetic spin-orbit interaction~\cite{chen2019non}, the Zitterbewegung effect~\cite{chen2019non,polimeno2021experimental,lovett2023observation,nalitov2015spin}, and non-Abelian Aharonov--Bohm interference~\cite{yang2019synthesis,chen2019non}. 
However, most of these photonic studies on non-Abelian gauge fields are in stationary systems, whereas they are much less explored in Floquet systems.

Floquet systems, of which Hamiltonians are periodic in time, can host topological phases as in their stationary counterparts~\cite{kitagawa2012topological,harper2020topology}. 
Moreover, an exotic topological property of Floquet systems is that they also support anomalous topological boundary modes~\cite{rudner2013anomalous}, which have no counterpart in stationary systems. The topologically protected edge mode can exist despite the Chern number being trivial. Discrete-time quantum walk (DTQW), where periodically multiplying unitary operators can be viewed as a stroboscopic simulation of time evolution by an effective Hamiltonian~\cite{asboth2012symmetries,kitagawa2012topological,kitagawa2010exploring}, can be arguably viewed as the simplest Floquet systems.
The Floquet phases in DTQW have been extensively studied, with the anomalous edge modes also appearing in DTQW~\cite{chen2018observation,asboth2015edge,groh2016robustness}. 
Moreover, synthetic gauge fields have been discussed in topological DTQW as well~\cite{chalabi2019synthetic,sajid2019creating,boada2017quantum}, but they are limited to Abelian ones. 
It thus calls for the exploration of the topological consequences of non-Abelian gauge fields in DTQW and, more broadly, in Floquet systems.

In this paper, we introduce non-Abelian gauge fields in time-multiplexed topological DTQW and study their topological consequences. We find that non-Abelian gauge fields can lead to Peierls substitution in both momenta and quasienergy, level repulsion, and spin texture exchanges in the Floquet bands of quantum walks. We propose topological walks in one and two dimensions and show that non-Abelian gauge fields can effectively manipulate their topological invariants. Non-Abelian gauge fields can also conveniently help simulate entangled walkers together with coin operations in different subspaces.

We introduce polarization-multiplexed photonic mesh lattice as a versatile platform for studying quantum walks immersed in non-Abelian gauge fields. Its Abelian version, featuring single-mode fibers~\cite{regensburger2011photon} or free-space optics, has been studied extensively and led to the creation of parity-time synthetic photonic lattices~\cite{regensburger2012parity}, the observation of parity-time-symmetric solitons~\cite{wimmer2015observation} and topological edge states~\cite{xiao2017observation,chen2018observation}, simulation of correlated quantum walks~\cite{schreiber20122d}, quantum walk in Abelian gauge fields~\cite{chalabi2019synthetic}, light funneling~\cite{weidemann2020topological} and so on.

\begin{figure}
	\centering
	\includegraphics[width=1\linewidth]{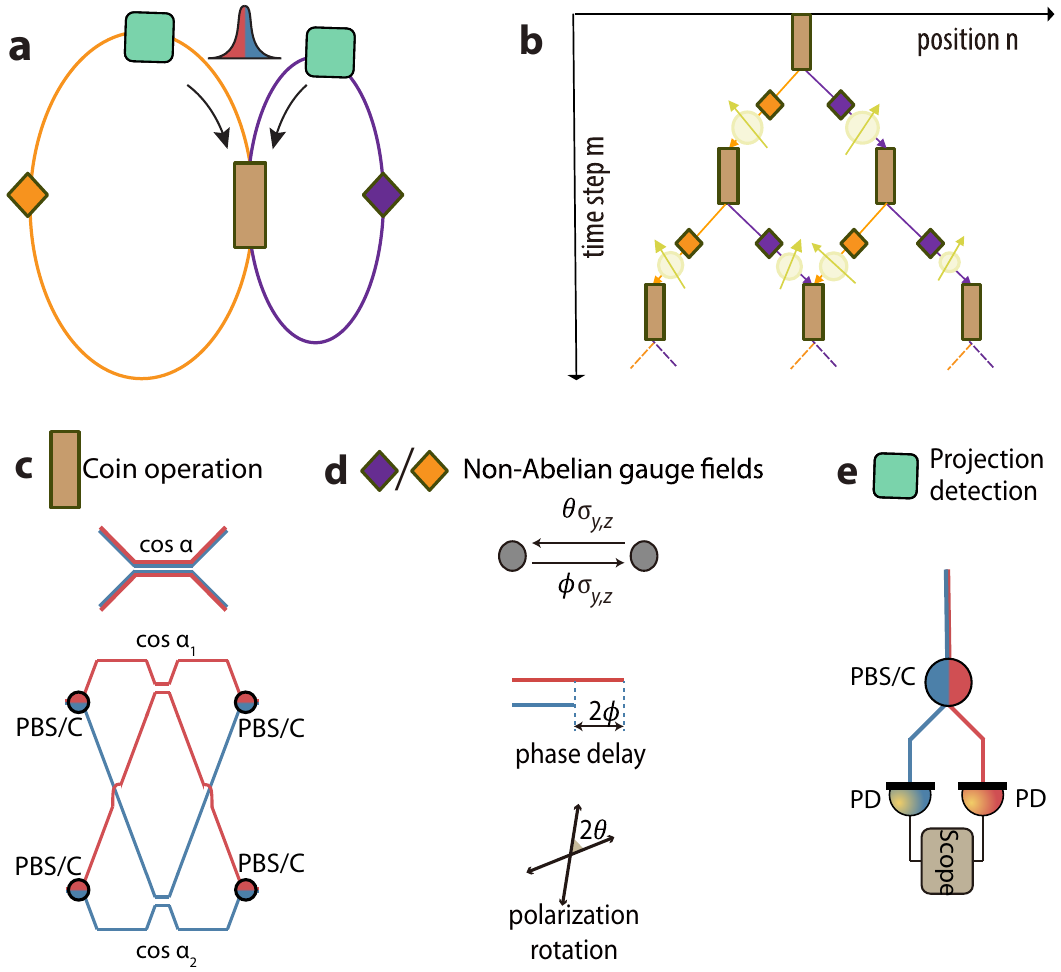}
	 \caption{%
	 	\textbf{Building blocks of synthetic non-Abelian gauge fields in polarization-multiplexed photonic mesh lattices. } 
	 \textbf{a.} Conceptual schematic. A light pulse (red and blue indicate polarization multiplexing) propagates in two coupled delay lines, giving rise to a four-dimensional Hilbert space. The brown rectangle is the beam splitter (c), the two diamonds are non-Abelian gauge fields (d), and the green square is polarization projection detection (e). 
     \textbf{b.} Equivalent time-bin encoded beam splitter network. The pulse is marked by the circle. When the pulse passes the beam splitter, it splits into two pulses that propagate in different fiber loops and experience different gauge fields (orange and purple diamond). The lattice is spanned by the discrete time step $m$ and discrete position $n$. Each lattice site $(m,n)$ is arrival-time encoded. 
     \textbf{c.} Coin operator implemented via a PM coupler with a split ratio characterized by $\cos{\alpha}$ (top). Red and blue lines represent horizontal and vertical polarization, which can have different coin angles (bottom).  
     \textbf{d.} Gauge fields $\e^{\iu \phi\sigmaz}$ and $\e^{\iu \theta\sigmay}$ coupling to opposite momenta (top) can be implemented by phase delay (middle) and polarization rotation (bottom). 
     \textbf{e.} Projection detection (PD) can be used for measuring the arrival time of the light pulse in different polarizations.
	 	}
	\label{fig1:schematic}
\end{figure}

Now, we consider a similar but generalized setup composed of polarization-maintaining (PM) elements. We will show below that this rationale could enable a convenient and scalable platform for quantum walks with a four-dimensional (4D) coin that is amenable to the incorporation and manipulation of synthetic non-Abelian gauge fields.
The generalization to a 4D coin is necessary for studying quantum walks in non-Abelian gauge fields~\cite{bisio2016quantum} because the spin and the coin space are identical in conventional quantum walk, and only Abelian gauge fields can only be encoded. 
We note that 4D quantum walk has been studied in~\cite{lorz2019photonic,schreiber20122d} but non-Abelian gauge fields have not been incorporated, and their topological consequences have not been discussed.
The conceptual schematic of our proposal setup is shown in Fig.~\ref{fig1:schematic}a.
We identify the horizontal and vertical polarization as the pseudospin $\sigma=\left\{\uparrow, \downarrow\right\}$. Meanwhile, the two coupled fiber loops $\tau=\left\{a,b\right\}$ become the coin space. Therefore, a 4D Hilbert space $\tau\otimes\sigma$ can be formed by the product between the coins and the pseudospins. 

The quantum walk can be realized with time-multiplexed bins in two coupled fiber loops, where a pulse round-trip within the two loops corresponds to left- and right-movers, respectively, on the synthetic one-dimensional chain.
In Fig.~\ref{fig1:schematic}b, the propagation of the light pulse inside the loop simulates the dynamics of a pulse in a mesh lattice made of many beam splitters (brown rectangles in Fig.~\ref{fig1:schematic}b).
One complete propagation of an optical pulse around the loop is equivalent to a walker hopping to the nearby site and experiencing different non-Abelian gauge fields (orange and purple diamonds in Fig.~\ref{fig1:schematic}b) coupled to opposite momenta.

On this platform, the two unequal long polarization-maintaining fiber loops are coupled by a polarization-maintaining (PM) coupler, which serves as the 2D coin operator. 
The 4D coin operator can be implemented via a PM coupler with a split ratio characterized by $\cos\alpha$ as shown in Fig.~\ref{fig1:schematic}c top, which assumes an equal split ratio $\cos\alpha$ for the two polarizations. 
If needed, the coin operator can be made polarization-dependent by replacing the single PM coupler with a composite structure made of two couplers and four polarization-maintaining beam splitters/combiners (PBS/Cs) (Fig.~\ref{fig1:schematic}c bottom) such that the two polarizations have two independent coin angles $(\alpha_1, \alpha_2)$.

Because the two coupled loops already separate the momenta towards opposite directions (Fig.~\ref{fig1:schematic}d top), synthetic non-Abelian gauge fields acting on the polarization state can be readily achieved by either phase delay (Fig.~\ref{fig1:schematic}d middle) or polarization rotation (Fig.~\ref{fig1:schematic}d bottom), corresponding to $\sigmaz$- and $\sigmay$-basis non-Abelian gauge fields, respectively.
The probability distribution of the quantum walk at each time step can be detected from the polarization projection detection (Fig.~\ref{fig1:schematic}e) that can be placed within both delay loops such that one can directly measure the pulse evolution within the entire 4D Hilbert space.

We first consider a one-dimensional (1D) quantum walk in non-Abelian gauge fields.
The conditional translational operating on the coin space $\tau$ is given by 
\begin{align}\label{eq:1Dtranslational}
T_x^\tau &= \sum_{x} \ket{x-1}\bra{x}\otimes \ket{a}\bra{a} +  \ket{x+1}\bra{x}\otimes \ket{b}\bra{b}
\end{align}
where the coin-dependent translation acts on the delay-loop space $\tau$ such that we can insert non-Abelian gauge fields onto the pseudospin space $\sigma$.
We first insert commutative SU(2) gauge fields into two loops. In momentum space, the walk protocol is given by 
\begin{align}
    U(k) = S(k)\cdot G(\theta\sigmay,\phi\sigmay)\cdot R(\alpha),
    \label{eq:4D_abelian walk}
\end{align}
where $\tau(\sigma)_{x,y,z}$ are Pauli matrices, $\tau_0/\sigma_0$ is an identity matrix, $R(\alpha) = \e^{\iu \alpha\tau_y}\otimes \sigma_0$ is the coin rotation operator, $S(k) = \e^{-\iu k \tau_z} \otimes \sigma_0$ is the conditional translation operator.
$G(\theta\sigmay,\phi\sigmay)\equiv \e^{\iu\theta\sigmay}\oplus\e^{\iu\phi\sigmay}$ contains the synthetic gauge fields for opposite directions in each coupled loop. 

The quantum walk in Eq.~\eqref{eq:4D_abelian walk} permits unitary block diagonalization (see Sec.~S1). Therefore, we can decouple the walker into two subspaces, where the dispersion relations are obtained as (see Sec.~S1) 
\begin{align}\label{eq:4D_abelian_energy}
    \cos\left(\epsilon\pm\delta\epsilon\right)=\cos\left(k\pm\delta k\right)\cos\alpha,
\end{align}
where $\delta k=(\theta-\phi)/2$ and $\delta\epsilon = (\theta+\phi)/2$ are the Peierls substitutions in momenta $k$ and quasienergy $\epsilon$, respectively. 
Evidently, when the gauge fields are absent $\theta=\phi=0$, the quintessential quantum-walk dispersion is restored as $\cos\epsilon = \cos k\cos \alpha$~\cite{kitagawa2010exploring}. 
Therefore, Eq.~\eqref{eq:4D_abelian_energy} shows that synthetic gauge fields can lead to Peierls substitution in both momenta and quasienergy in Floquet Hermitian systems, whereas the substitution only applies to momenta in stationary problems. We further studied non-Abelian quantum walk featuring $G(\theta\sigmay,\phi\sigmaz)$ in Eq.~\eqref{eq:4D_abelian walk} (see Sec.~S2), which displays avoided crossing and spin texture exchanges between different Floquet bands in a similar fashion as those in spin-orbit coupling.

We next show that non-Abelian gauge fields can be used to realize topological quantum walk and topological phase transitions. We first provide a DTQW on a 1D chain:
\comment{yy}{this walker is inspired by some past work, right? Where is the reference?}\comment{ZHP}{I cannot find the origin of the equation from Omar's notes.}
\comment{yy}{I will ask Omar}
\begin{equation}\label{eq:1D_topological_walk}
U(k) = S(k)R(\alpha_2) G_2(\phi) R(\alpha_2)S(k)R(\alpha_1) G_1(\theta) R(\alpha_1),
\end{equation}
where $S(k) = \e^{-\iu k\tau_z}\otimes \sigma_0$,
$G_1(\theta) = \e^{\iu\theta\sigma_y}\oplus e^{-\iu\theta\sigma_y}$,
$G_2(\phi)= \e^{\iu\phi \sigma_z}\oplus e^{-\iu\phi\sigma_z}$,
and
$R(\alpha_i) = \e^{-\iu\alpha_i\tau_y}\otimes \sigma_0$.
To derive the bulk topological invariants, we perform the analysis proposed by Asb\'{o}th and Obuse~\cite{asboth2013bulk}. We transform $U(k)$ into the time-symmetric frame, where $U'(k) = SU(k)S^{\dagger}$ and $S = G_1(\theta/2)R(\alpha_1)$. In the time-symmetric frame, we can find the chiral operator $\Gamma' = \tau_x \otimes \sigma_0$ such that $\Gamma'\hat{U'}(k)\Gamma'^{\dagger} = \hat{U'}^{\dagger}$. In the time-symmetric frame, we can have two chiral symmetric Floquet operators
$U' = FG$
and
$U'' = GF$
where $F = G_1(\theta/2)R(\alpha_1)S(k)R(\alpha_2) G_2(\phi/2)$ and $G = G_2(\phi/2) R(\alpha_2)S(k)R(\alpha_1) G_1(\theta/2)$. The winding number for each chiral-symmetric Floquet operator can be defined as follows: let $H'(k) = [U'(k)^{\dagger} - U'(k)]/2\iu$, in a chiral basis, $H'(k)$ is block--off-diagonal, the upper right block $h'(k)$ is used to define winding number $\nu'$
\begin{equation}\label{eq:4D_windingnumber}
\nu' = \frac{1}{2\pi\iu}\int_{-\pi}^{\pi}dk\frac{d}{dk}\ln\mathrm{det}h'(k)
\end{equation}
and $\nu''$ can be defined analogously for $U^{''}$.
Thus, the bulk topological invariants $(\nu_0, \nu_{\pi})$ can be constructed from winding number of $U'$ and $U''$ via
\begin{equation}\label{eq:4D_topological_invariant}
(\nu_0, \nu_{\pi}) = \left(\frac{\nu'+\nu''}{2},\frac{\nu'-\nu''}{2}\right).
\end{equation}

\begin{figure}[htbp]
	\centering
	\includegraphics[width=1\linewidth]{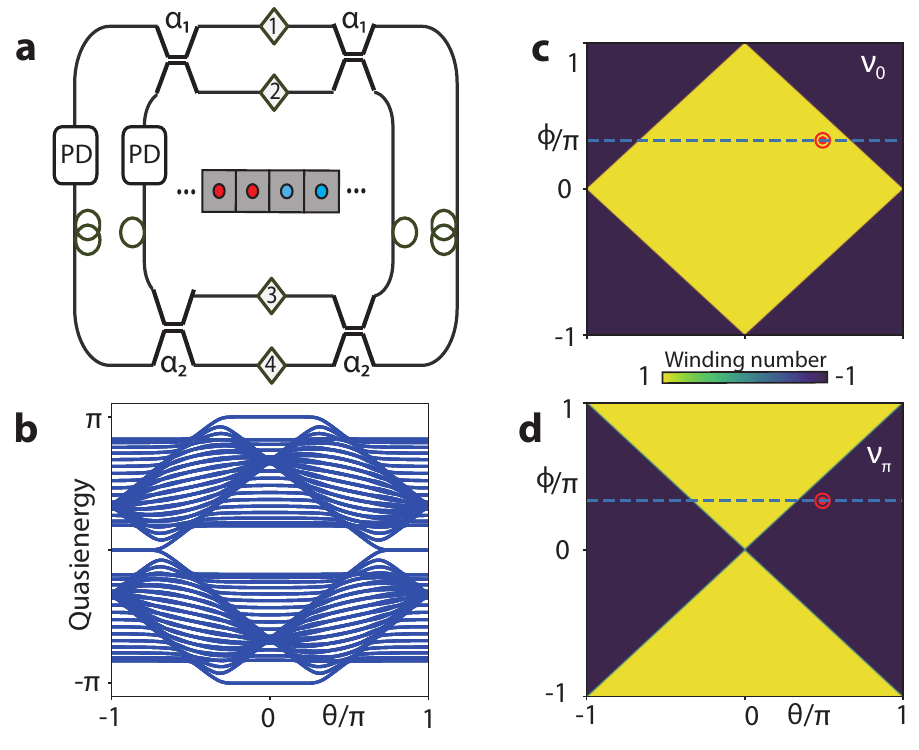}
	 \caption{%
	 	\textbf{1D topological quantum walk in non-Abelian gauge fields. }
	 	\textbf{a.} Setup for 1D topological quantum walk. Diamonds 1-4 represents gauge fields $\e^{\iu \theta \sigmay}$, $\e^{-\iu \theta \sigmay}$, $\e^{-\iu \phi \sigmaz}$ and $\e^{\iu \phi \sigmaz}$ respectively. PD stands for projected detection. %
        \textbf{b.} Domain wall quasienergy spectrum form by red point and blue dashed line in c-d. 
        \textbf{c-d.} Phase diagram induced by gauge fields $(\theta,\phi)$ for zero and $\pi$ mode. For the blue dashed line, $\phi = \pi/3$ and $\theta$ from $-\pi$ to $\pi$. For the red point, $\theta = 1.55, \phi = \pi/3.$  We fix $\alpha_1 = \pi/4, \alpha_2 = \pi/4$ throughout.
	 	}
	\label{fig3:1Dtopo}
\end{figure}

In Fig.~\ref{fig3:1Dtopo}a, we propose an experiment setup for the walker in Eq.~\eqref{eq:1D_topological_walk} based on the optical elements in Fig.~\ref{fig1:schematic}.
In this setup, four beam splitters realize the coin operators $\alpha_1$ and $\alpha_2$. They connect four delay lines encoding the split-step translation in one dimension. Meanwhile, the beam splitters also connect four diamond-shape gauge-field components to realize $G_1(\theta)$ and $G_2(\phi)$, where the amplitude of $\theta$ and $\phi$ becomes time-dependent to realize a domain wall between two bulks (red and blue domains in the middle inset of Fig.~\ref{fig3:1Dtopo}a; also see further discussion of this setup presented in Sec.~S3). 

This domain wall can be used to confirm the bulk analysis above and the bulk-boundary correspondence. In particular, it can showcase the topological phase transition enabled by non-Abelian gauge fields.
In this domain-wall $(\theta,\phi)$ configuration, both left and right bulks have fixed $\phi = \pi/3$, and the left bulk has a fixed $\theta = 1.55$ (red dots in Fig.~\ref{fig3:1Dtopo}c and d).
Meanwhile, $\theta$ varies from $-\pi$ to $\pi$ for the right bulk (dashed line in Fig.~\ref{fig3:1Dtopo}c and d). 
The quasienergy spectrum of this domain wall as a function of the varying $\theta$ is plotted in Fig.~\ref{fig3:1Dtopo}b, where the appearance of zero and $\pi$ modes exhibits consistency with the calculation of bulk invariants. Specifically, the appearance of edge modes can be explained by $(\nu_0,\nu_\pi)$ phase diagram as shown in Fig.~\ref{fig3:1Dtopo}c-d, where zero ($\pi$) modes appear when $\nu_0$ ($\nu_\pi$) differs in the left and right bulk. It is thus evident that non-Abelian gauge fields $(\theta,\phi)$ can lead to Floquet topological phase transitions. 

\begin{figure*}
	\centering
	\includegraphics[width=1\linewidth]{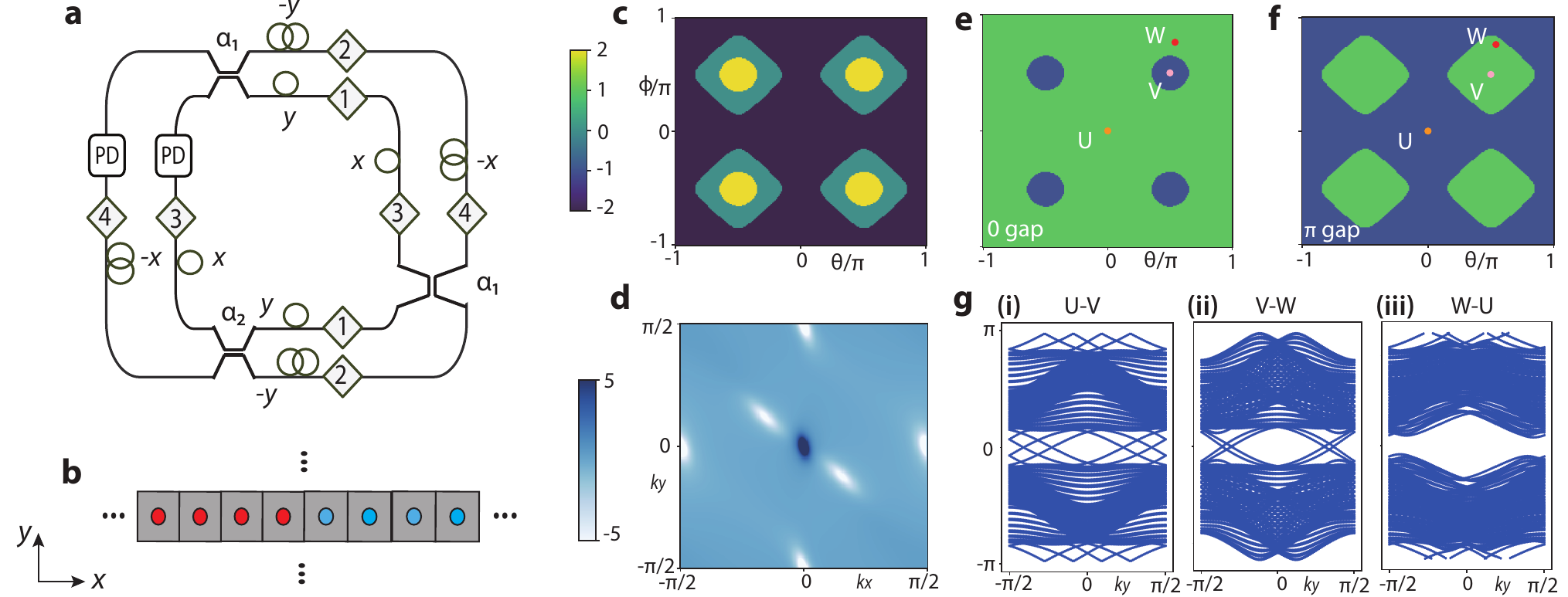}
	 \caption{%
	 	\textbf{Proposed setup, Chern and RLBL phase diagrams, and quasienergy spectra of 2D topological quantum walk.}
	 	\textbf{a. }Setup for a 2D topological quantum walk in non-Abelian gauge fields.
            Diamond 1-4 represents gauge fields $\e^{-\iu \phi \sigmay}$,$\e^{\iu \phi \sigmay}$,$\e^{-\iu \theta \sigmaz}$ and $\e^{\iu \theta \sigmaz}$ respectively. PD stands for projection detection.
            \textbf{b.} The geometry of the domain wall. It is a periodic domain wall made of two bulks in the $x$-direction, while the $y$ direction is continuously translational invariant.
	 	\textbf{c.} Chern number summed over the top two energy bands in $(\theta,\phi)$ space.
        \textbf{d. }Berry curvature at point W.
        \textbf{e-f.} RLBL phase diagrams of zero gap and $\pi$ gap induced by gauge fields $(\theta,\phi)$. Topological phases are distinguished by RLBL invariants $W_{0,\pi} = \pm 1$. $(\theta,\phi) = (0,0)$, $(1.7, 2.4)$, and $(\pi/2,\pi/2)$ for points U, V, and W, respectively. 
        \textbf{g.} 2D domain-wall spectra formed by (i) U-V, (ii) V-W, and (iii) W-U. The appearance and disappearance of boundary states are consistent with the RLBL phase diagrams.
        Here we fix coin angles $\alpha_1 = \pi/6, \alpha_2 = \pi/6$ throughout.
	 	}
	\label{fig4:2Dtopo}
\end{figure*}

Next, we extend our discussion to non-Abelian gauge fields in 2D topological DTQW featuring anomalous topological boundary modes. 
We construct a 2D DTQW consisting of spin-dependent translations dressed by non-Abelian gauge fields and separated by coin-flip operations:
\begin{align}
    U_{\mathrm{2D}}(\mathbf{k}) = S_x(k_x)R_2S_y(k_y)R_1S_x(k_x)S_y(k_y)R_1,
    \label{eq:2D_non-abelian}
\end{align}
where 
\begin{align}
\begin{split}
    & S_{x,y}(k_{x,y}) =
    \left(
    \begin{array}{cc}
         \e^{-\iu k_{x,y}}U_{x,y} & 0  \\
         0& \e^{\iu k_{x,y}}U^{\dagger}_{x,y}
    \end{array}
    \right),\\
    & R_{1,2} = \e^{-\iu\alpha_{1,2}\tau_y} \otimes \sigma_0,    U_x(\theta) = \e^{\iu \theta \sigmaz},    U_y(\phi) = \e^{\iu \phi \sigmay }.
\end{split}
\end{align}
The Floquet operator is inspired by Ref.~\cite{kitagawa2012topological} but generalized with non-Abelian gauge fields.
Here, $U_{x,y}$ are the minimally-coupled non-Abelian gauge fields since $S_{x,y} = \e^{-\iu \tau_z \otimes (k_{x,y}\sigma_0-A_{x,y})}$, where $\textbf{\textit{A}} = (\theta \sigmaz, \phi \sigmay)$ is spatially homogeneous but still exhibits nonzero curvature because of the non-commutativity of the different Cartesian components. 
This 2D quantum walk can be equally realized on a photonic setup as shown in Fig.~\ref{fig4:2Dtopo}a, which consists of three polarization-maintaining beam splitters with their ports connected to fibers of different lengths such that they map the $\pm x$ and $\pm y$ directions to different time delays.   

This walker is topologically nontrivial, as indicated by the Chern number phase diagram in $(\theta,\phi)$ space shown in Fig.~\ref{fig4:2Dtopo}c and an example Berry curvature in Fig.~\ref{fig4:2Dtopo}d. 
the Berry curvature satisfies $\Omega(k_x,k_y) = \Omega(-k_x,-k_y)$; it follows from the inversion symmetry that $PU_{\mathrm{2D}}(k_x,k_y)P^{\dagger} = U_{\mathrm{2D}}(-k_x,-k_y)$ where the inversion operator $P = \tau_y \otimes \sigmax$. 
The full variation of the gauge fields between $(-\pi,\pi)$ can lead to Chern number variations among integers $\left\{0,\pm 2 \right\}$ for half of the Floquet bands. 
Nevertheless, the Chern number alone is insufficient to explain the anomalous Floquet boundary modes. 

We continue to study this topological phase transition induced by the non-Abelian gauge fields based on the so-called Rudner-Lindner-Berg-Levin (RLBL) invariant~\cite{sajid2019creating,rudner2013anomalous} that is unique to Floquet systems.
To calculate the RLBL invariant for our quantum walk, we map our walker to a time-dependent Hamiltonian~\cite{asboth2015edge}. Specifically, we use a non-overlapping sequence of pulses where at any time, only one type of pulse is switched on; the time-dependent Hamiltonian can thus be constructed from a non-overlapping sequence of pulses (details in Sec.~S4). 
Fig.~\ref{fig4:2Dtopo}e-f shows the zero-gap and $\pi$-gap phase diagram of the RLBL invariant, quantized at $\pm 1$, as a function of non-Abelian gauge fields $(\theta, \phi)$, indicating that the RLBL invariant can be effectively tuned by non-Abelian gauge fields.
The RLBL invariant calculation is also corroborated by an alternative method of magnetic quantum walk~\cite{asboth2017spectral} described in Sec.~S5). 
Incidentally, their phase diagrams in Fig.~\ref{fig4:2Dtopo}c and e-f 
confirm the relation between RLBL invariant and Chern number $W[U_{\pi}]-W[U_{0}] = C_{0,\pi}$~\cite{rudner2013anomalous}, where $C_{0,\pi}$ denotes the sum of the Chern numbers of all
Floquet bands that lie in between 0 and $\pi$, and $W_{\pi(0)}$ is the $\pi(\mathrm{zero})$-gap invariant. 
Fig.~\ref{fig4:2Dtopo}g(i-iii) shows the band structure of a geometry containing periodic domain walls made of two bulks in the $x$-direction, while the $y$ direction is continuously translational invariant (Fig.~\ref{fig4:2Dtopo}b). If the Floquet operators of the two bulks have distinct or the same $W_{\pi(0)}$, edge states are present or absent in the $\pi$ (zero) gap as shown in Fig.~\ref{fig4:2Dtopo}g (i-iii). Note that the number of edge modes in Fig.~\ref{fig4:2Dtopo}g is twice as many as the predictions from the RLBL invariants. This is because the RLBL invariant can be calculated in a reduced BZ $(-\pi/2,\pi/2]$ because of the split-step walker, whereas in the domain-wall calculation $k_x$ effectively sample the entire BZ $(-\pi,\pi]$.

\begin{figure}
	\centering
	\includegraphics[width=1\linewidth]{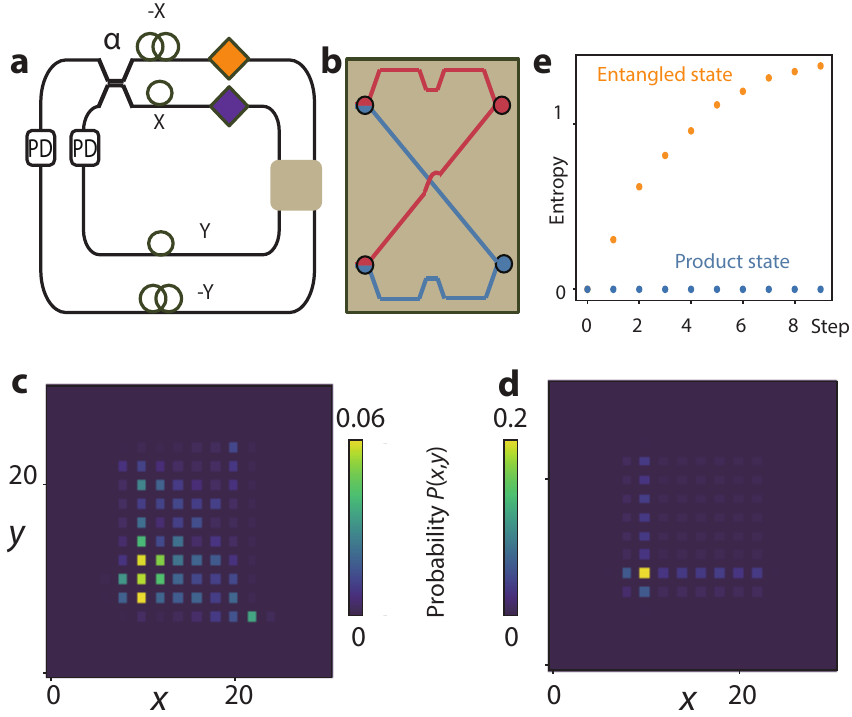}
	 \caption{%
	 	\textbf{Simulation of entangled walker with non-Abelian gauge fields}
	 	\textbf{a.} Setup for simulating entangled quantum walk where the two diamonds are non-Abelian gauge fields. The gray box is explained in b. 
        \textbf{b.} Optical realization for coin operation in pseudo-spin space. It couples different pseudospins to different loops via two PBS/Cs and two PM couplers. 
        \textbf{c-d.}
        Simulated probability distribution $P(x,y)$ after 9 steps of entangled and non-entangled 2D quantum walk with an initial state $\ket{a}\otimes\ket{\uparrow}$. The walker is launched at $(x=15,y=15)$ in the middle of the lattice. Gauge field components are different and the same in c and d, respectively.
        \textbf{e.} Evolution of von Neumann entropy for entangled and non-entangled states. The entangled state is enabled by the gauge fields $\e^{-\iu \theta \sigma_y}$ and $\e^{-\iu \phi \sigma_{z}}$ where we choose $\theta = \phi = \pi/4$. 
	 	}
	\label{fig5:entanglement}
\end{figure}

We next extend our discussion to the simulation of entangled walkers.
It has been established that a single walker in a 2D lattice can be used to simulate two entangled walkers on a 1D lattice~\cite{schreiber20122d}.
Concretely, a spatial distribution from a 2D lattice with positions $(x_1,x_2)$ can be interpreted as a coincidence measurement for two walkers at position $x_1$ and $x_2$ propagating on the same linear graph. 
Fig.~\ref{fig5:entanglement}a shows the setup for an example 2D quantum walk, which contains two types of coin operators whose condition translation in the $x$ and $y$ directions are determined by the spatial loop subspace and the pseudospin subspace, respectively. The Floquet operator is explicitly given by
\begin{align}\label{2DEntanglementFloquet}
    U = T_y^\sigma U_{yz}(\theta,\phi)T_x^\tau R(\alpha).
\end{align}
where $R(\alpha) = \e^{-\iu\alpha\tau_y}$ is the coin operator and $U_{yz}(\theta,\phi)=\e^{-\iu \theta \sigma_y} \oplus \e^{-\iu \phi \sigma_z}$ are the non-Abelian gauge fields. 
Note that the conditional translation operators now operate on different subspaces: \begin{align}\label{eq:2DEntanglement}
T_y^\sigma &= \sum_{y} \ket{y-1}\bra{y}\otimes \ket{\uparrow}\bra{\uparrow} +  \ket{y+1}\bra{y}\otimes \ket{\downarrow}\bra{\downarrow}
\end{align}
operates in the pseudospin subspace (can be realized by the structure shown in Fig.~\ref{fig5:entanglement}b), and $T_x^\tau$ is defined in Eq.~\eqref{eq:1Dtranslational} (which operates on the loop subspace in all the rest parts of the paper).

The entanglement simulation is induced by the distinct block diagonal elements in $U_{yz}(\theta,\phi)$. 
Specifically, $U_{yz}(\theta,\phi)$ can be written as conditional gate operation $U_{yz}(\theta,\phi) = \begin{pmatrix}
1 & 0 \\
0 & 0
\end{pmatrix}\otimes \e^{-\iu\theta \sigmay} + \begin{pmatrix}
0 & 0 \\
0 & 1
\end{pmatrix}\otimes \e^{-\iu\phi \sigmaz}$, meaning that $\e^{-\iu\theta \sigmay}$ or $\e^{-\iu\phi \sigmaz}$ is applied to the pseudospin space if the coin state is $\ket{a}$ or $\ket{b}$, respectively. 
We can illustrate this with a few examples.
If we choose $\theta=0,\phi = \pi/2$, the conditional gate operation $U_{yz}(\theta,\phi)$ becomes the controlled-Z gate with an additional phase $\e^{\iu \pi/2}$; similarly, If we choose $\theta = \pi/2, \phi = 0$, the conditional gate operation becomes the controlled-Y gate; the controlled Hadamard gate can be implemented by choosing $\theta = \pi/4$ and $\phi = 0$.   
As a result, these non-Abelian gauge fields could cause entanglement between the loop subspace and the pseudospin subspace, thus creating correction between positions $(x_1,x_2)$. 
In contrast, if the block diagonal elements in $U_{yz}(\theta,\phi)$ become equal to each other, no entanglement can be generated, and the probability should permit decomposition into two independent quantum walks.
Fig.~\ref{fig5:entanglement}c-d show the probability $P(x_1,x_2)$ for entangled and non-entangled quantum walks after nine steps for an initial state $\ket{a}\otimes\ket{\uparrow}$. For the non-entangled quantum walk in Fig.~\ref{fig5:entanglement}d, the probability can be decomposed into two independent 1D quantum walks $P(x_1,x_2) = P(x_1)P(x_2)$; such decomposition is not possible for the entangled case in Fig.~\ref{fig5:entanglement}c. 
In Fig.~\ref{fig5:entanglement}e, we calculate the von Neumann entropy against the evolution step. The entropy increases with the number of evolution steps for the entangled quantum walk $U_{yz}(\theta,\phi)$ and remains zero for the non-entangled walk.  

In conclusion, we theoretically propose a concrete and versatile photonic scheme, which is based on polarization-multiplexed photonic mesh lattices, for realizing quantum walks immersed in non-Abelian gauge fields.
Based on this platform, we introduce examples of one- and two-dimensional quantum walks whose topology and entanglement simulations can be effectively manipulated with non-Abelian gauge fields. 
The proposed setup is friendly to fiber optics and free-space optics realizations and could give rise to opportunities in creating and controlling photonic Floquet phases.

\emph{Acknowledgments.}
We thank Ali Ghorashi, Jinbing Hu, Mark Rudner, Yannick Salamin and Bengy Tsz Tsun Wong for helpful discussions.



\bibliographystyle{apsrev4-1-fixspace} 

%

\end{document}


\title{Supplementary Information\\ Topological quantum walk in synthetic non-Abelian gauge fields}


\author{Zehai Pang}
\affiliation{\hkuaffil}
\author{Omar Abdelghani}
\affiliation{\mitaffil}
\author{Marin~Solja\v{c}i\'{c}}
\affiliation{\mitaffil}
\author{Yi~Yang}
\email{yiyg@hku.hk}
\affiliation{\hkuaffil}

\maketitle

\setlength{\parindent}{0em}
\setlength{\parskip}{.5em}

\noindent{\textbf{\textsf{CONTENTS}}}\\ 
\twocolumngrid
\begingroup 
    \let\bfseries\relax 
    \deactivateaddvspace 
    \tableofcontents
\endgroup
\onecolumngrid

\hyphenation{eigen-index}

\section{Peierls substitution in momentum and quasienergy}\label{S1}

In this section, we address a simple four-dimensional (4D) quantum walk (featuring commutative SU(2) gauge fields) which is block-diagonalizable from its two particle-hole symmetry operators. We then obtain the analytical expression for the dispersion relation showing the explicit Peierls substitution in momentum and quasienergy. 

\subsection{Symmetry analysis}
We consider the 4D walker
\begin{align}
U(k) &= S(k)\cdot G(\theta\sigmay,\phi\sigmay)\cdot R(\alpha)
\label{smeq:abelian_walk}
\end{align}
where
\begin{align}
G(\theta\sigmay,\phi\sigmay) &=
    \left(
    \begin{array}{cc}
         \e^{\iu \theta \sigmay} & 0  \\
         0& \e^{\iu \phi \sigmay}
    \end{array}
    \right).
\end{align}
Two particle-hole operators are identified in this quantum walk
\begin{align}
    \mathcal{C}_1 &= \tau_0\otimes\sigmai K\\
    \mathcal{C}_2 &= \tau_0\otimes \iu\sigmay K
\end{align}
such that
\begin{align}
    \mathcal{C}U(k)\mathcal{C}^\dagger = U(-k).
\end{align}
Thus, the composite unitary operator $O = \mathcal{C}_1 \mathcal{C}_2 = \sigmai\otimes \iu\sigmay$ is a unitary operator that satisfies
\begin{align}
    OU(k)O^\dagger=U(k).
\end{align}
and thus can block diagonalize $U(k)$ with $O$'s eigenvectors $V_O$ such that
\begin{align}\label{Eq:Diagonalizd quantum walk}
    V_O^\dagger O V_O &= \mathrm{diag}(-\iu,-\iu,\iu,\iu),\\
    V_O^\dagger U(k) V_O &= \tilde{U}(k)=\left(
    \begin{array}{cc}
    \tilde{U}_a(k) & 0 \\
    0 & \tilde{U}_b(k)
    \end{array}
    \right)
\end{align}
where
\begin{align}
    \tilde{U}_a(k)&=\left(
    \begin{array}{cc}
    \e^{-\iu(k+\theta)}\cos\alpha & -e^{-i (k+\theta)}\sin\alpha \\
    e^{i (k-\phi )}\sin\alpha & e^{i (k-\phi )}\cos\alpha 
    \end{array}
    \right),\\
    \tilde{U}_b(k)&=\left(
        \begin{array}{cc}
        \e^{-i (k-\theta )}\cos\alpha & -e^{-i (k-\theta )}\sin\alpha \\
        \e^{\iu(k+\phi)}\sin\alpha & \e^{\iu(k+\phi)}\cos\alpha \\
    \end{array}
    \right).
\end{align}

\subsection{Eigenvalue expression of the block-diagonalizable 4D walker}

We first consider $\tilde{U}_a(k)$. We define $\delta=(\theta-\phi)/2$, $\tilde{k}=k+\delta$ and $\sigma = (\theta+\phi)/2$ and obtain
\begin{align}
\tilde{U}_a(k)&=\left(
    \begin{array}{cc}
 \e^{-\iu(\tilde{k}+\sigma)}\cos\alpha & -e^{-i (\tilde{k}+\sigma)}\sin\alpha \\
 e^{i (\tilde{k}-\sigma)}\sin\alpha & e^{i (\tilde{k}-\sigma )}\cos\alpha 
 \end{array}
 \right)
\end{align}
The quasienergy of the walker satisfies
\begin{align}
    \mathrm{Det}(\tilde{U}_a(k)-\lambda I)=0\\
    \lambda^2-\left[\e^{-\iu(\tilde{k}+\sigma)}+\e^{\iu(\tilde{k}-\sigma)}\right]\cos\alpha\lambda +\e^{-2\iu\sigma}=0\\
    \left(\lambda\e^{-\iu\sigma}\right)^2-2\cos\tilde{k}\cos\alpha\cdot\lambda\e^{-\iu\sigma} +1=0.    
    \label{eq:4D_a_dispersion}
\end{align}
Here $\lambda = \e^{-\iu \varepsilon}$, $\varepsilon$ is the quasienergy. Take the real part of \eqref{eq:4D_a_dispersion}, we have the dispersion relation
\begin{align}
    \cos\left(\epsilon+\sigma\right)=\cos\left(k+\delta\right)\cos\alpha.
\end{align}

Analogously, $\tilde{U}_b(k)$ becomes:
\begin{align}
\tilde{U}_b(k) &=\left(
    \begin{array}{cc}
 \e^{-\iu(\dbtilde{k}-\sigma)}\cos\alpha & -e^{-i (\dbtilde{k}-\sigma)}\sin\alpha \\
 e^{i (\dbtilde{k}+\sigma)}\sin\alpha & e^{i (\dbtilde{k}+\sigma )}\cos\alpha 
 \end{array}
 \right)
\end{align}

where $\dbtilde{k}=k-\delta$. The quasienergy of the walker satisfies
\begin{align}
    \mathrm{Det}(\tilde{U}_b(k)-\lambda I)=0\\
    \lambda^2-\left[\e^{-\iu(\dbtilde{k}-\sigma)}+\e^{\iu(\dbtilde{k}+\sigma)}\right]\cos\alpha\lambda +\e^{2\iu\sigma}=0\\
    \left(\lambda\e^{\iu\sigma}\right)^2-2\cos\dbtilde{k}\cos\alpha\cdot\lambda\e^{\iu\sigma} +1=0.    
    \label{eq:4D_b_dispersion}
\end{align}
Similarly, we have the dispersion relation
\begin{align}
    \cos\left(\epsilon-\sigma\right)=\cos\left(k-\delta\right)\cos\alpha.
\end{align}
Therefore, the quasienergy dispersion relation of Eq.\eqref{smeq:abelian_walk} is given by 
\begin{align}\label{eq:4D_abelian_energy}
    \cos\left(\epsilon\pm\delta\epsilon\right)=\cos\left(k\pm\delta k\right)\cos\alpha,
\end{align}
where $\delta k=(\theta-\phi)/2$ and $\delta\epsilon = (\phi+\theta)/2$.

\section{Minimal 1D quantum walk in non-Abelian gauge fields}\label{S2}

We consider a minimal quantum walk featuring non-commutative SU(2) gauge fields
\begin{align}
U(k) &= S(k)\cdot G(\theta\sigmay,\phi\sigmaz)\cdot R(\alpha)
\label{smeq:min_NA_walk}
\end{align}
Where the walker experiences non-commuting gauge fields when it moves to the left and to the right. By symmetry analysis, for $\theta,\phi \notin \{0,\pi\}$,  the particle-hole operator is given by $\mathcal{C} = \tau_0 \otimes \iu \sigmay K$, time-reversal operator $\mathcal{T} = -\tau_x \otimes \frac{1}{\sqrt{2}}(\iu\sigma_0 \mp \sigmax) K$ for $\theta = \pm \phi$, and chiral operator $\mathcal{S} = \mathcal{T}\cdot\mathcal{C}$.

\begin{figure}[h]
	\centering
	\includegraphics[width=1\linewidth]{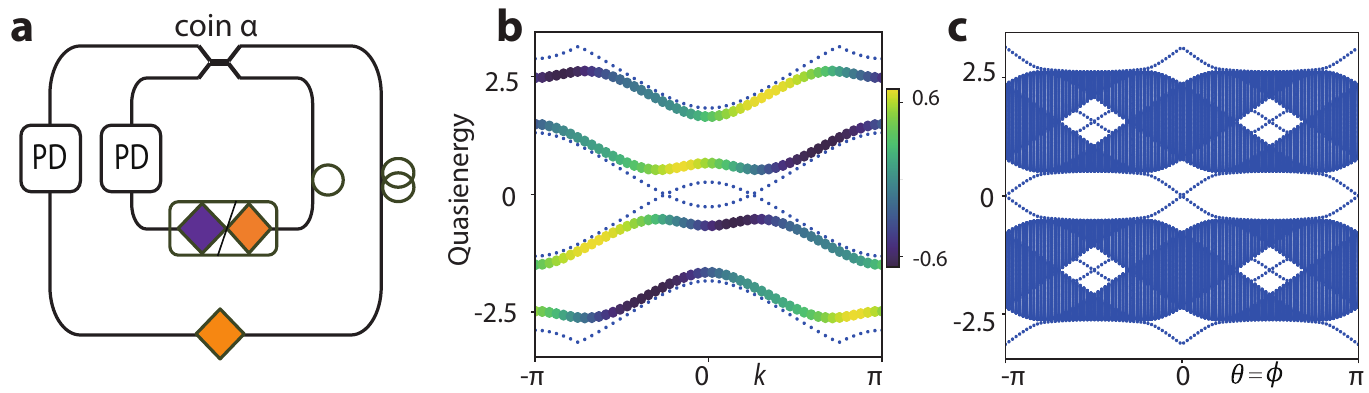}
	 \caption{%
	 	\textbf{Minimal one-dimensional quantum walk in non-Abelian gauge fields.} \textbf{a.} Setup for a minimal quantum walk in non-Abelian gauge fields in one dimension. PD stands for projected detection. The SU(2) gauge fields in the two loops can be either commutative [both gauge field elements in orange; Eq.\eqref{smeq:abelian_walk}) or non-commutative [two gauge field elements in orange and purple, respectively; Eq.\eqref{smeq:min_NA_walk}]. The two loops of unequal length are connected by a polarization-maintaining coupler with a split ratio characterized by $\cos{\alpha}$.  
	   \textbf{b.} Quasienergy bands of walker Eq.\eqref{smeq:abelian_walk} (blue dashed line) and walker Eq.\eqref{smeq:min_NA_walk} (colorful line). When the gauge fields become non-commutative, the Floquet bands show level repulsion and spin texture exchange. The spin orientation is obtained for the basis $\sigmaz\otimes\sigmaz$. Here, we choose $\alpha = \pi/4, \theta =\phi= \pi/3.$ 
        \textbf{c.} Quasienergy spectra under reflecting boundary condition, with $\alpha = \pi/4, \theta =\phi.$
	 	}
	\label{figS:coupling}
\end{figure}

\subsection{Experimental setup}
In Fig.~\ref{figS:coupling}a, we propose an experiment setup for realizing both walkers Eq.~\eqref{smeq:abelian_walk} and Eq.~\eqref{smeq:min_NA_walk}. In the setup, non-Abelian gauge fields are implemented by polarization rotator and phase retarder, respectively (See Fig.~1 in the main text). 
For the quantum walk with commutative SU(2) gauge fields, the quasienergy spectrum is indicated by the blue dashed lines in Fig.~\ref{figS:coupling}b, which clearly demonstrate the Peierls substitution in both momentum and quasienergy. 
Non-Abelian gauge fields in Eq.~\eqref{smeq:min_NA_walk} can lift the degeneracies and change the quasienergy spectrum from level crossing to level repulsion as shown by the colored bands (where color encodes spin texture) in Fig.~\ref{figS:coupling}b. It displays avoided crossing and spin texture exchanges between different Floquet bands in a similar fashion as those in spin-orbit coupling. Moreover, the degeneracies of the zero and $\pi$ modes will be lifted under non-zero gauge fields  as shown in Fig.~\ref{figS:coupling}c.

\subsection{Symmetry analysis}

\subsubsection{Chiral Symmetry}
The chiral operator is defined by 
\begin{align}\label{eq:4D_chiraloperator}
\mathcal{S} U \mathcal{S} = U^{\dagger}
\end{align}
where $\mathcal{S}$ is the chiral operator. To find the chiral operator for Eq.~\eqref{smeq:min_NA_walk}, we need to first gauge transform the quantum walk in a time-symmetric frame, $U' = R(\alpha/2)S(k)GR(\alpha/2)$. For the chiral symmetry, it requires
\begin{align}\label{eq:4D_chiralSymmetry}
\begin{split}
\mathcal{S} U' \mathcal{S} &= U'^{\dagger}\\
\mathcal{S} R(\alpha/2)S(k)GR(\alpha/2) \mathcal{S} &= R(-\alpha/2)S(-k)G^{-1}R(-\alpha/2)
\end{split}
\end{align}
Here, we take the ansatz that $\mathcal{S} = \sigmax \otimes M$, where $M^2 = I$. To find exact form of $M$, we can substitute $\mathcal{S}$ into Eq.~\eqref{eq:4D_chiralSymmetry} and have
\begin{align}
\begin{split}
M(\e^{\iu\theta\sigmay} + \e^{\iu\phi\sigmaz})M = \e^{-\iu\theta\sigmay} + \e^{-\iu\phi\sigmaz}\\
M(\e^{\iu\theta\sigmay} - \e^{\iu\phi\sigmaz})M = -\e^{-\iu\theta\sigmay} + \e^{-\iu\phi\sigmaz}
\end{split}
\end{align}
The above equations imply $M\e^{\iu\theta\sigmay}M = \e^{-\iu\phi\sigmaz}$. $M$ can also be decomposed into the Pauli basis where $M = a\sigmax + b\sigmay +c\sigmaz$ and $a^2 + b^2 + c^2 = 1$. Therefore
\begin{align}
\begin{split}
M\sigmay M &= \begin{pmatrix}
\frac{\Delta-\iu \sin{\phi}}{\iu \sin{\theta}} & 0 \\
0 & \frac{\Delta + \iu \sin{\phi}}{\iu \sin{\theta}}
\end{pmatrix},\\
M\sigmay M &= 2ab\sigmax - a^2\sigmay + (b^2-c^2)\sigmay + 2bc\sigmaz
\end{split}
\end{align}
Here $\Delta = \cos{\phi} - \cos{\theta}$. By matching coefficients, it requires $a = 0$ and $\abs{b}=\abs{c}$. For the chiral symmetry to exist, it requires $\Delta = 0$. Therefore, the chiral operator in the symmetric frame is given by
\begin{align}
\begin{split}
\mathcal{S} &= \tau_x \otimes \frac{1}{\sqrt{2}}(\sigmay-\sigmaz),\quad \text{for $\theta = \phi$}.\\
\mathcal{S} &= \tau_x \otimes \frac{1}{\sqrt{2}}(\sigmay+\sigmaz),\quad \text{for $\theta = -\phi$}.
\end{split}
\end{align}
Therefore, the chiral symmetry exists only when $\abs{\theta} = \abs{\phi}$. 

\subsubsection{Particle-hole Symmetry}
The particle-hole symmetry $\mathcal{C}$ is defined as 
\begin{align}
\mathcal{C} U(k) \mathcal{C}^{-1} = U(-k).
\end{align}
In the symmetric frame, we can see particle-hole operator is in the form of $\mathcal{C} = \tau_0 \otimes \iu\sigmay K$, where $K$ is the complex conjugation operator. 

\subsubsection{Time-reversal Symmetry}
For the time-reversal symmetry $\mathcal{T}$, it is defined as 
\begin{align}
\mathcal{T} U(k) \mathcal{T}^{-1} = U^{-1}(-k)
\end{align}
$\mathcal{T}$ is the time-reversal operator. The time-reversal operator can also be obtained from relation $\mathcal{S} = \mathcal{T} \cdot \mathcal{C}$. 
Therefore, $\mathcal{T} = \mathcal{S} \cdot \mathcal{C}^{-1} = -\tau_x \otimes \frac{1}{\sqrt{2}}(\iu \sigma_0 \pm \sigmax) K$ for $\theta = \mp \phi$. 

\section{Edge excitation in one-dimensional topological quantum walk in non-Abelian gauge fields}\label{S3}

\begin{figure}[H]
	\centering
	\includegraphics[width=0.5\linewidth]{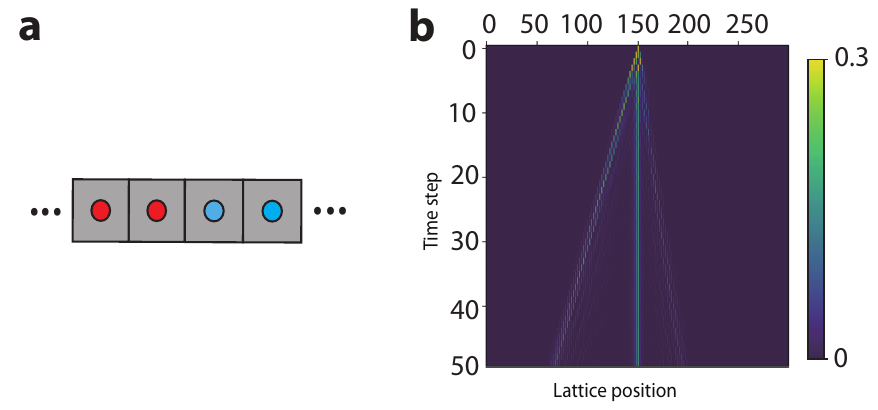}
	 \caption{%
	 	\textbf{Edge mode excitation in the 1D topological quantum walk in non-Abelian gauge fields.}
            \textbf{a.} Configuration of the domain wall.
	 	\textbf{b. }Numerical probability distribution $P(m,n)$ as a function of time step $(m)$ and lattice position $(n)$ for the inhomogeneous 1D DTQW with a domain wall at $n=150$. For $n < 150$ (left bulk), $\theta = 1.55, \phi = \pi/3.$ and for $n > 150$ (right bulk), $\theta = 0.5, \phi = \pi/3.$ Left and right bulk have different $\nu_0$, indicating the existence of edge states. 
        The walker is launched at $n= 150$ with initial state $\ket{b}\otimes\ket{\uparrow}$. $\alpha_1 = \pi/4, \alpha_2 = -\pi/4$.
	 	}
	\label{figS:1Dtopo}
\end{figure}

We show the numerical calculation of probability distribution $P(m,n)$ as a function of time step $(m)$ and lattice position $(n)$ showing the excitation of topological boundary mode with a domain-wall configuration. The domain wall configuration is shown in Fig.~\ref{figS:1Dtopo}a, it shows a geometry containing periodic domain walls made of two bulks in the $x$-direction. The 1D topological quantum walk is given by Eq.~(4) in the main text. 
Here, we consider a domain-wall configuration where left and right bulks have different topological invariants. In Fig.~\ref{figS:1Dtopo}b, the excitation is mainly trapped in the edge of the domain wall, which indicates the existence of edge states. Note that the excitation is given by the walker is initialized in the middle $(n=150)$ with state $\ket{b}\otimes\ket{\uparrow}$, which couple to both edge and bulk modes. Only the edge mode will be excited if we prepare excitation with the eigenstate of the edge mode.

\section{Two-dimensional topological quantum walk in non-Abelian gauge fields}\label{S4}

In this section, we first provide a detailed derivation showing how our proposed setups can faithfully realize the target quantum walkers in non-Abelian gauge fields. After that, we construct a time-dependent Hamiltonian and show how to obtain the topological invariant of the walker. We also discuss the spectral flow of the walker.

\begin{figure}[h]
	\centering
	\includegraphics[width=0.8\linewidth]{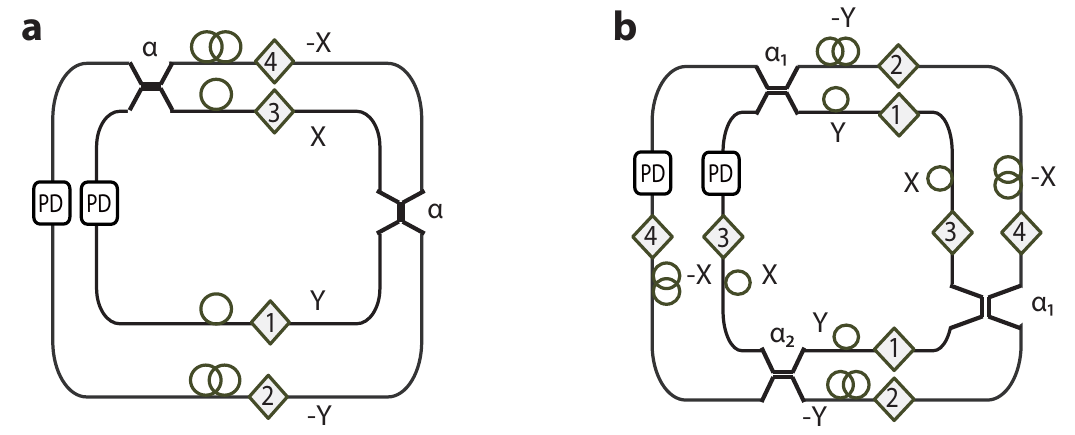}
	 \caption{%
	 	\textbf{Experimental setups for 2D quantum walk in non-Abelian gauge fields.}
            \textbf{a.} Setup for a minimal 2D quantum walk in non-Abelian gauge fields $\mathbf{A}=(\theta\sigmaz,\phi\sigmay,0)$. \textbf{b.} Setup for a 2D topological quantum walk in non-Abelian gauge fields.
            Diamond 1-4 represents gauge fields $\e^{-\iu \phi \sigmay}$,$\e^{\iu \phi \sigmay}$,$\e^{-\iu \theta \sigmaz}$ and $\e^{\iu \theta \sigmaz}$ respectively. PD stands for projected detection. Gauge fields can be implemented by phase retarders and polarization rotators.
	 	}
	\label{figS:2Dtopo}
\end{figure}

\subsection{Optical implementation and derivation}

We first show a detailed transfer matrix derivation of how to realize a 2D quantum walk in the simplest spatially-homogeneous non-Abelian gauge fields $\mathbf{A}=(\theta\sigmaz,\phi\sigmay,0)$.

The quantum walk is given by 
\begin{align}
    U(\mathbf{k}) = G_y(\phi)S_y(k_y)R(\alpha)G_x(\theta)S_x(k_x)R(\alpha),
    \label{eq:goldman_walker}
\end{align}
where
$R(\alpha)=R_0(\alpha)\otimes\sigma_0$ is the coin operator,
$S_{x,y}=\exp(\iu k_{x,y})\sigmai\oplus\exp(-\iu k_{x,y})\sigmai$ are the step operator,
$G_x(\theta)=\exp(\iu\theta\sigmaz)\oplus\exp(-\iu\theta\sigmaz)$ is the gauge field along $x$ direction, 
and
$G_y(\phi)=\exp(\iu\phi\sigmay)\oplus\exp(-\iu\phi\sigmay)$ is the gauge field along the $y$ direction. 

The associated optical setup for Eq.~\eqref{eq:goldman_walker} is shown in Fig.~\ref{figS:2Dtopo}a. 
The 4D coin operator is the 50:50 coupler given by
\begin{align}
    R_{y}(\alpha)=R_{0,y}(\alpha)\otimes\sigma_0=
    \frac{1}{\sqrt{2}}
    \left(
    \begin{array}{cccc}
         1 & 0 & -1 & 0 \\
         0 & 1 & 0 & -1 \\
         1 & 0 & 1 & 0 \\
         0 & 1 & 0 & 1         
    \end{array}
    \right),
\end{align}
where the Hilbert space is $(a_\uparrow,a_\downarrow,b_\uparrow,b_\downarrow)$ ($a$ and $b$ denotes outside and inside loop in Fig.~1 in main text, respectively). 

In the following, the evolution of pulses given by Eq.~\eqref{eq:goldman_walker} in the two-dimensional quantum walk is analyzed (notations below similar to those in the Supplementary section of Ref.~\cite{chalabi2019synthetic}). Two pulses just before the first beam splitter, $\begin{pmatrix}
U^n_{x,y}\\
D^n_{x,y}
\end{pmatrix}$, at time step $n$ will produce output pulses as move in the $x$ direction:
\begin{align}
    \left(
    \begin{array}{c}
    W_{x-1,y}\\
    K_{x+1,y}
    \end{array}
    \right)
    =
    \frac{1}{\sqrt{2}}
    \left(
    \begin{array}{cc}
    \e^{-\iu\theta\sigmaz} & 0\\
    0 & \e^{\iu\theta\sigmaz} 
    \end{array}
    \right)
    \left(
    \begin{array}{cccc}
         1 & 0 & -1 & 0 \\
         0 & 1 & 0 & -1 \\
         1 & 0 & 1 & 0 \\
         0 & 1 & 0 & 1         
    \end{array}
    \right)
    \left(
    \begin{array}{c}
    U^n_{x,y}\\
    D^n_{x,y}
    \end{array}
    \right),
\end{align}
where $\begin{pmatrix}
U^n_{x,y}\\
D^n_{x,y}
\end{pmatrix}$ is the initial state for the pulse at time step $n$ on lattice site $(x,y)$ and
$\begin{pmatrix}
W_{x-1,y}\\
K_{x+1,y}
\end{pmatrix}$ is the state after evolution in $x$ direction.
Similarly, in the $y$ direction, $\begin{pmatrix}
W_{x,y}\\
K_{x,y}
\end{pmatrix}$ will produce output pulse. The evolution is
\begin{align}
    \left(
    \begin{array}{c}
    U^{n+1}_{x,y-1}\\
    D^{n+1}_{x,y+1}
    \end{array}
    \right)
    =
    \frac{1}{\sqrt{2}}
    \left(
    \begin{array}{cc}
    \e^{-\iu\phi\sigmay} & 0\\
    0 & \e^{\iu\phi\sigmay} 
    \end{array}
    \right)
    \left(
    \begin{array}{cccc}
         1 & 0 & -1 & 0 \\
         0 & 1 & 0 & -1 \\
         1 & 0 & 1 & 0 \\
         0 & 1 & 0 & 1         
    \end{array}
    \right)
    \left(
    \begin{array}{c}
    W_{x,y}\\
    K_{x,y}
    \end{array}
    \right),
\end{align}
Consider the pulse of $\left(
    \begin{array}{c}
    W_{x-1,y}\\
    0
    \end{array}
    \right)$ excited by $\left(
    \begin{array}{c}
    U^n_{x,y}\\
    D^n_{x,y}
    \end{array}
    \right)$:
\begin{align}
        \left(
    \begin{array}{c}
    U^{n+1}_{x-1,y-1}\\
    D^{n+1}_{x-1,y+1}
    \end{array}
    \right)
    =\frac{1}{\sqrt{2}}
    \left(
    \begin{array}{cc}
    \e^{-\iu\phi\sigmay} & 0\\
    0 & \e^{\iu\phi\sigmay} 
    \end{array}
    \right)
    \left(
    \begin{array}{cccc}
         1 & 0 & -1 & 0 \\
         0 & 1 & 0 & -1 \\
         1 & 0 & 1 & 0 \\
         0 & 1 & 0 & 1         
    \end{array}
    \right)
    \left(
    \begin{array}{c}
    W_{x-1,y}\\
    0
    \end{array}
    \right),\\
    =
    \frac{1}{\sqrt{2}}
    \left(
    \begin{array}{cc}
    \e^{-\iu\phi\sigmay} & 0\\
    0 & \e^{\iu\phi\sigmay} 
    \end{array}
    \right)
    \left(
    \begin{array}{cccc}
         1 & 0 & -1 & 0 \\
         0 & 1 & 0 & -1 \\
         1 & 0 & 1 & 0 \\
         0 & 1 & 0 & 1         
    \end{array}
    \right)
    \left(
    \begin{array}{cccc}
         1 & 0 & 0 & 0 \\
         0 & 1 & 0 & 0 \\
         0 & 0 & 0 & 0 \\
         0 & 0 & 0 & 0         
    \end{array}
    \right)
    \frac{1}{\sqrt{2}}
    \left(
    \begin{array}{cc}
    \e^{-\iu\theta\sigmaz} & 0\\
    0 & \e^{\iu\theta\sigmaz} 
    \end{array}
    \right)
    \left(
    \begin{array}{cccc}
         1 & 0 & -1 & 0 \\
         0 & 1 & 0 & -1 \\
         1 & 0 & 1 & 0 \\
         0 & 1 & 0 & 1         
    \end{array}
    \right)
    \left(
    \begin{array}{c}
    U^n_{x,y}\\
    D^n_{x,y}
    \end{array}
    \right),\\
    =
    \frac{1}{2}
    \left(
    \begin{array}{cccc}
     \e^{-\iu \theta } \cos{\phi} & -\e^{\iu \theta } \sin{\phi} & -\e^{-\iu \theta } \cos{\phi} & \e^{\iu \theta } \sin{\phi} \\
     \e^{-\iu \theta } \sin{\phi} & \e^{\iu \theta } \cos{\phi} & -\e^{-\iu \theta } \sin{\phi} & -\e^{\iu \theta } \cos{\phi} \\
     \e^{-\iu \theta} \cos{\phi} & \e^{\iu \theta} \sin{\phi} & -\e^{-\iu \theta} \cos{\phi} & -\e^{\iu \theta} \sin{\phi} \\
     -\e^{-\iu \theta} \sin{\phi} & \e^{\iu \theta} \cos{\phi} & \e^{-\iu \theta} \sin{\phi} & -\e^{\iu \theta} \cos{\phi} \\
    \end{array}
    \right)
    \left(
    \begin{array}{c}
    U^n_{x,y}\\
    D^n_{x,y}
    \end{array}
    \right),
\end{align}
Similarly, Consider the pulse of $\left(
    \begin{array}{c}
    0\\
    K_{x+1,y}
    \end{array}
    \right)$ excited by $\left(
    \begin{array}{c}
    U^n_{x,y}\\
    D^n_{x,y}
    \end{array}
    \right)$:
\begin{align}
        \left(
    \begin{array}{c}
    U^{n+1}_{x+1,y-1}\\
    D^{n+1}_{x+1,y+1}
    \end{array}
    \right)
    =\frac{1}{\sqrt{2}}
    \left(
    \begin{array}{cc}
    \e^{-\iu\phi\sigmay} & 0\\
    0 & \e^{\iu\phi\sigmay} 
    \end{array}
    \right)
    \left(
    \begin{array}{cccc}
         1 & 0 & -1 & 0 \\
         0 & 1 & 0 & -1 \\
         1 & 0 & 1 & 0 \\
         0 & 1 & 0 & 1         
    \end{array}
    \right)
    \left(
    \begin{array}{c}
    0\\
    K_{x+1,y}
    \end{array}
    \right),\\
    =
    \frac{1}{\sqrt{2}}
    \left(
    \begin{array}{cc}
    \e^{-\iu\phi\sigmay} & 0\\
    0 & \e^{\iu\phi\sigmay} 
    \end{array}
    \right)
    \left(
    \begin{array}{cccc}
         1 & 0 & -1 & 0 \\
         0 & 1 & 0 & -1 \\
         1 & 0 & 1 & 0 \\
         0 & 1 & 0 & 1         
    \end{array}
    \right)
    \left(
    \begin{array}{cccc}
         0 & 0 & 0 & 0 \\
         0 & 0 & 0 & 0 \\
         0 & 0 & 1 & 0 \\
         0 & 0 & 0 & 1         
    \end{array}
    \right)
    \frac{1}{\sqrt{2}}
    \left(
    \begin{array}{cc}
    \e^{-\iu\theta\sigmaz} & 0\\
    0 & \e^{\iu\theta\sigmaz} 
    \end{array}
    \right)
    \left(
    \begin{array}{cccc}
         1 & 0 & -1 & 0 \\
         0 & 1 & 0 & -1 \\
         1 & 0 & 1 & 0 \\
         0 & 1 & 0 & 1         
    \end{array}
    \right)
    \left(
    \begin{array}{c}
    U^n_{x,y}\\
    D^n_{x,y}
    \end{array}
    \right)\\
    =
    \frac{1}{2}
    \left(
    \begin{array}{cccc}
     -\e^{\iu \theta} \cos{\phi} & \e^{-\iu \theta} \sin{\phi} & -\e^{-\iu \theta} \cos{\phi} & \e^{-\iu \theta} \sin{\phi} \\
     -\e^{\iu \theta} \sin{\phi} & -\e^{-\iu \theta} \cos{\phi} & -\e^{\iu \theta} \sin{\phi} & -\e^{-\iu \theta} \cos{\phi} \\
     \e^{\iu \theta} \cos{\phi} & \e^{-\iu \theta} \sin{\phi} & \e^{\iu \theta} \cos{\phi} & \e^{-\iu \theta} \sin{\phi} \\
     -\e^{\iu \theta} \sin{\phi} & \e^{-\iu \theta} \cos{\phi} & -\e^{\iu \theta} \sin{\phi} & \e^{-\iu \theta} \cos{\phi} \\
    \end{array}
    \right)
    \left(
    \begin{array}{c}
    U^n_{x,y}\\
    D^n_{x,y}
    \end{array}
    \right).    
\end{align}
Taken together,
\begin{align}
\begin{split}
    U^{n+1}_{x,y}
    &=\frac{1}{2}
    \left(
    \begin{array}{cccc}
     \e^{-\iu \theta} \cos{\phi} & -\e^{\iu \theta} \sin{\phi} & -\e^{-\iu \theta} \cos{\phi} & \e^{\iu \theta} \sin{\phi} \\
     \e^{-\iu \theta} \sin{\phi} & \e^{\iu \theta} \cos{\phi} & -\e^{-\iu \theta} \sin{\phi} & -\e^{\iu \theta} \cos{\phi} 
    \end{array}
    \right)
    \left(
    \begin{array}{c}
    U^n_{x+1,y+1}\\
    D^n_{x+1,y+1}
    \end{array}
    \right)\\
    &+
    \frac{1}{2}
    \left(
    \begin{array}{cccc}
     -\e^{\iu \theta} \cos{\phi} & \e^{-\iu \theta} \sin{\phi} & -\e^{\iu \theta} \cos{\phi} & \e^{-\iu \theta} \sin{\phi} \\
     -\e^{\iu \theta} \sin{\phi} & -\e^{-\iu \theta} \cos{\phi} & -\e^{\iu \theta} \sin{\phi} & -\e^{-\iu \theta} \cos{\phi} 
    \end{array}
    \right)
    \left(
    \begin{array}{c}
    U^n_{x-1,y+1}\\
    D^n_{x-1,y+1}
    \end{array}
    \right).\\
    &=\frac{1}{2}
    \left(
    \begin{array}{cc}
     \e^{-\iu \theta} \cos{\phi} & -\e^{\iu \theta} \sin{\phi} \\
     \e^{-\iu \theta} \sin{\phi} & \e^{\iu \theta} \cos{\phi} 
    \end{array}
    \right)
    \left(
    \begin{array}{c}
    U^n_{x+1,y+1}-D^n_{x+1,y+1}
    \end{array}
    \right)\\
    &+
    \frac{1}{2}
    \left(
    \begin{array}{cccc}
     -\e^{\iu \theta} \cos{\phi} & \e^{-\iu \theta} \sin{\phi}  \\
     -\e^{\iu \theta} \sin{\phi} & -\e^{-\iu \theta} \cos{\phi} 
    \end{array}
    \right)
    \left(
    \begin{array}{c}
    U^n_{x-1,y+1}+D^n_{x-1,y+1}
    \end{array}
    \right).\\
\end{split}
\label{eq:Un}
\end{align}

\begin{align}
\begin{split}
    D^{n+1}_{x,y}
    &=\frac{1}{2}
    \left(
    \begin{array}{cccc}
     \e^{-\iu \theta} \cos{\phi} & \e^{\iu \theta} \sin{\phi} & -\e^{-\iu \theta} \cos{\phi} & -\e^{\iu \theta} \sin{\phi} \\
     -\e^{-\iu \theta} \sin{\phi} & \e^{\iu \theta} \cos{\phi} & \e^{-\iu \theta} \sin{\phi} & -\e^{\iu \theta} \cos{\phi} 
    \end{array}
    \right)
    \left(
    \begin{array}{c}
    U^n_{x+1,y-1}\\
    D^n_{x+1,y-1}
    \end{array}
    \right)\\
    &+
    \frac{1}{2}
    \left(
    \begin{array}{cccc}
     \e^{\iu \theta} \cos{\phi} & \e^{-\iu \theta} \sin{\phi} & \e^{\iu \theta} \cos{\phi} & \e^{-\iu \theta} \sin{\phi} \\
     -\e^{\iu \theta} \sin{\phi} & \e^{-\iu \theta} \cos{\phi} & -\e^{\iu \theta} \sin{\phi} & \e^{-\iu \theta} \cos{\phi} 
    \end{array}
    \right)
    \left(
    \begin{array}{c}
    U^n_{x-1,y-1}\\
    D^n_{x-1,y-1}
    \end{array}
    \right)\\
    &=\frac{1}{2}
    \left(
    \begin{array}{cccc}
     \e^{-\iu \theta} \cos{\phi} & \e^{\iu \theta} \sin{\phi} \\
     -\e^{-\iu \theta} \sin{\phi} & \e^{\iu \theta} \cos{\phi}
    \end{array}
    \right)
    \left(
    \begin{array}{c}
    U^n_{x+1,y-1}-D^n_{x+1,y-1}
    \end{array}
    \right)\\
    &+
    \frac{1}{2}
    \left(
    \begin{array}{cccc}
     \e^{\iu \theta} \cos{\phi} & \e^{-\iu \theta} \sin{\phi} \\
     -\e^{\iu \theta} \sin{\phi} & \e^{-\iu \theta} \cos{\phi} 
    \end{array}
    \right)
    \left(
    \begin{array}{c}
    U^n_{x-1,y-1}+D^n_{x-1,y-1}
    \end{array}
    \right).    
\end{split}
\label{eq:Dn}
\end{align}
To clearly see relation between evolution of pulse and Eq.~\eqref{eq:goldman_walker}, define
\begin{align}
    S^n_{x,y} = \frac{1}{\sqrt{2}}(U^n_{x,y}+D^n_{x,y})\\
    P^n_{x,y} = \frac{1}{\sqrt{2}}(U^n_{x,y}-D^n_{x,y}).
\end{align}
Thus
\begin{align}
    S^{n+1}_{x,y} 
    &=
    \left(
    \begin{array}{cc}
     \e^{-\iu \theta} \cos{\phi} & -\e^{\iu \theta} \sin{\phi} \\
     \e^{-\iu \theta} \sin{\phi} & \e^{\iu \theta} \cos{\phi} 
    \end{array}
    \right)
    \left(
    \begin{array}{c}
    P^n_{x+1,y+1}
    \end{array}
    \right)
    +
    \left(
    \begin{array}{cccc}
     -\e^{\iu \theta} \cos{\phi} & \e^{-\iu \theta} \sin{\phi}  \\
     -\e^{\iu \theta} \sin{\phi} & -\e^{-\iu \theta} \cos{\phi} 
    \end{array}
    \right)
    \left(
    \begin{array}{c}
    S^n_{x-1,y+1}
    \end{array}
    \right)\\
    &+
    \left(
    \begin{array}{cccc}
     \e^{-\iu \theta} \cos{\phi} & \e^{\iu \theta} \sin{\phi} \\
     -\e^{-\iu \theta} \sin{\phi} & \e^{\iu \theta} \cos{\phi}
    \end{array}
    \right)
    \left(
    \begin{array}{c}
    P^n_{x+1,y-1}
    \end{array}
    \right)
    +
    \left(
    \begin{array}{cccc}
     \e^{\iu \theta} \cos{\phi} & \e^{-\iu \theta} \sin{\phi} \\
     -\e^{\iu \theta} \sin{\phi} & \e^{-\iu \theta} \cos{\phi} 
    \end{array}
    \right)
    \left(
    \begin{array}{c}
    S^n_{x-1,y-1}
    \end{array}
    \right).   \\     
    &=\e^{\iu k_x+\iu k_y}
    \left(
    \begin{array}{cc}
     \e^{-\iu \theta} \cos{\phi} & -\e^{\iu \theta} \sin{\phi} \\
     \e^{-\iu \theta} \sin{\phi} & \e^{\iu \theta} \cos{\phi} 
    \end{array}
    \right)
    \left(
    \begin{array}{c}
    P^n_{x,y}
    \end{array}
    \right)
    +
    \e^{-\iu k_x+\iu k_y}
    \left(
    \begin{array}{cc}
     -\e^{\iu \theta} \cos{\phi} & \e^{-\iu \theta} \sin{\phi}  \\
     -\e^{\iu \theta} \sin{\phi} & -\e^{-\iu \theta} \cos{\phi} 
    \end{array}
    \right)
    \left(
    \begin{array}{c}
    S^n_{x,y}
    \end{array}
    \right)\\
    &+
    \e^{\iu k_x -\iu k_y}
    \left(
    \begin{array}{cc}
     \e^{-\iu \theta} \cos{\phi} & \e^{\iu \theta} \sin{\phi} \\
     -\e^{-\iu \theta} \sin{\phi} & \e^{\iu \theta} \cos{\phi}
    \end{array}
    \right)
    \left(
    \begin{array}{c}
    P^n_{x,y}
    \end{array}
    \right)
    +
    \e^{-\iu k_x -\iu k_y}
    \left(
    \begin{array}{cc}
     \e^{\iu \theta} \cos{\phi} & \e^{-\iu \theta} \sin{\phi} \\
     -\e^{\iu \theta} \sin{\phi} & \e^{-\iu \theta} \cos{\phi} 
    \end{array}
    \right)
    \left(
    \begin{array}{c}
    S^n_{x,y}
    \end{array}
    \right).        
\end{align}
\begin{align}
    P^{n+1}_{x,y} 
    &=
    \left(
    \begin{array}{cc}
     \e^{-\iu \theta} \cos{\phi} & -\e^{\iu \theta} \sin{\phi} \\
     \e^{-\iu \theta} \sin{\phi} & \e^{\iu \theta} \cos{\phi} 
    \end{array}
    \right)
    \left(
    \begin{array}{c}
    P^n_{x+1,y+1}
    \end{array}
    \right)
    +
    \left(
    \begin{array}{cccc}
     -\e^{\iu \theta} \cos{\phi} & \e^{-\iu \theta} \sin{\phi}  \\
     -\e^{\iu \theta} \sin{\phi} & -\e^{-\iu \theta} \cos{\phi} 
    \end{array}
    \right)
    \left(
    \begin{array}{c}
    S^n_{x-1,y+1}
    \end{array}
    \right)\\
    &-
    \left(
    \begin{array}{cccc}
     \e^{-\iu \theta} \cos{\phi} & \e^{\iu \theta} \sin{\phi} \\
     -\e^{-\iu \theta} \sin{\phi} & \e^{\iu \theta} \cos{\phi}
    \end{array}
    \right)
    \left(
    \begin{array}{c}
    P^n_{x+1,y-1}
    \end{array}
    \right)
    -
    \left(
    \begin{array}{cccc}
     \e^{\iu \theta} \cos{\phi} & \e^{-\iu \theta} \sin{\phi} \\
     -\e^{\iu \theta} \sin{\phi} & \e^{-\iu \theta} \cos{\phi} 
    \end{array}
    \right)
    \left(
    \begin{array}{c}
    S^n_{x-1,y-1}
    \end{array}
    \right).   \\     
    &=\e^{\iu k_x+\iu k_y}
    \left(
    \begin{array}{cc}
     \e^{-\iu \theta} \cos{\phi} & -\e^{\iu \theta} \sin{\phi} \\
     \e^{-\iu \theta} \sin{\phi} & \e^{\iu \theta} \cos{\phi} 
    \end{array}
    \right)
    \left(
    \begin{array}{c}
    P^n_{x,y}
    \end{array}
    \right)
    +
    \e^{-\iu k_x+\iu k_y}
    \left(
    \begin{array}{cc}
     -\e^{\iu \theta} \cos{\phi} & \e^{-\iu \theta} \sin{\phi}  \\
     -\e^{\iu \theta} \sin{\phi} & -\e^{-\iu \theta} \cos{\phi} 
    \end{array}
    \right)
    \left(
    \begin{array}{c}
    S^n_{x,y}
    \end{array}
    \right)\\
    &-
    \e^{\iu k_x -\iu k_y}
    \left(
    \begin{array}{cc}
     \e^{-\iu \theta} \cos{\phi} & \e^{\iu \theta} \sin{\phi} \\
     -\e^{-\iu \theta} \sin{\phi} & \e^{\iu \theta} \cos{\phi}
    \end{array}
    \right)
    \left(
    \begin{array}{c}
    P^n_{x,y}
    \end{array}
    \right)
    -
    \e^{-\iu k_x -\iu k_y}
    \left(
    \begin{array}{cc}
     \e^{\iu \theta} \cos{\phi} & \e^{-\iu \theta} \sin{\phi} \\
     -\e^{\iu \theta} \sin{\phi} & \e^{-\iu \theta} \cos{\phi} 
    \end{array}
    \right)
    \left(
    \begin{array}{c}
    S^n_{x,y}
    \end{array}
    \right).            
\end{align}
Jointly,
{
\begin{align}
    \left(
    \begin{array}{c}
    S^{n+1}_{x,y}\\
    P^{n+1}_{x,y}
    \end{array}
    \right)
    = \begin{pmatrix}
A & B \\
C & D
\end{pmatrix}\left(
    \begin{array}{c}
    S^{n}_{x,y}\\
    P^{n}_{x,y}
    \end{array}
    \right)
\end{align}
}
Where
\begin{equation}
\begin{split}    
A &=  \e^{-\iu k_x+\iu k_y}
    \begin{pmatrix}
     -\e^{\iu \theta} \cos{\phi} & \e^{-\iu \theta} \sin{\phi}  \\
     -\e^{\iu \theta} \sin{\phi} & -\e^{-\iu \theta} \cos{\phi} 
    \end{pmatrix}	
    +
    \e^{-\iu k_x -\iu k_y}
    \begin{pmatrix}
     \e^{\iu \theta} \cos{\phi} & \e^{-\iu \theta} \sin{\phi} \\
     -\e^{\iu \theta} \sin{\phi} & \e^{-\iu \theta} \cos{\phi}
     \end{pmatrix}\\
B &=  \e^{\iu k_x+\iu k_y}
    \begin{pmatrix}
     \e^{-\iu \theta} \cos{\phi} & -\e^{\iu \theta} \sin{\phi} \\
     \e^{-\iu \theta} \sin{\phi} & \e^{\iu \theta} \cos{\phi} 
   \end{pmatrix}
    +
    \e^{\iu k_x -\iu k_y}
    \begin{pmatrix}
     \e^{-\iu \theta} \cos{\phi} & \e^{\iu \theta} \sin{\phi} \\
     -\e^{-\iu \theta} \sin{\phi} & \e^{\iu \theta} \cos{\phi}
    \end{pmatrix}\\
C &=  \e^{\iu k_x+\iu k_y}
   \begin{pmatrix}
     \e^{-\iu \theta} \cos{\phi} & -\e^{\iu \theta} \sin{\phi} \\
     \e^{-\iu \theta} \sin{\phi} & \e^{\iu \theta} \cos{\phi} 
   \end{pmatrix}
    -
    \e^{\iu k_x -\iu k_y}
    \begin{pmatrix}
     \e^{-\iu \theta} \cos{\phi} & \e^{\iu \theta} \sin{\phi} \\
     -\e^{-\iu \theta} \sin{\phi} & \e^{\iu \theta} \cos{\phi}
    \end{pmatrix}\\
D &=  \e^{-\iu k_x+\iu k_y}
    \begin{pmatrix}
     -\e^{\iu \theta} \cos{\phi} & \e^{-\iu \theta} \sin{\phi}  \\
     -\e^{\iu \theta} \sin{\phi} & -\e^{-\iu \theta} \cos{\phi} 
   \end{pmatrix}
    -
    \e^{-\iu k_x -\iu k_y}
    \begin{pmatrix}
     \e^{\iu \theta} \cos{\phi} & \e^{-\iu \theta} \sin{\phi} \\
     -\e^{\iu \theta} \sin{\phi} & \e^{-\iu \theta} \cos{\phi} 
    \end{pmatrix}
\end{split}
\end{equation}

which evidently contains the evolution operator $U(\mathbf{k}) = (A \: B; C \: D)$.
One can show that this evolution operator is unitarily equivalent to Eq.~\eqref{eq:goldman_walker}.
The implementation of the walker Eq.\eqref{eq:goldman_walker} serves as a foundation for more complicated topological quantum walks in non-Abelian gauge fields.

Next, we generalize the above setup to the 2D topological quantum walk in the main text
\begin{align}
    U_{\mathrm{2D}}(k_x,k_y) = S_x(k_x)R_2S_y(k_y)R_1S_x(k_x)S_y(k_y)R_1,
    \label{eqS:2D_topolo}
\end{align}
By similar derivation, Eq.~\eqref{eqS:2D_topolo} can be realized in Fig.~\ref{figS:2Dtopo}b, which consists of three polarization-maintaining beam splitters with their ports connected to fibers of different lengths such that they map the $\pm x$ and $\pm y$ directions to different time delays. 
In contrast to the simple 2D walker [Eq.~\eqref{eq:goldman_walker}] that moves one step in $x$ and $y$  direction within one period time, Eq.~\eqref{eqS:2D_topolo} is a split-step walk; it moves two steps in $x$ and $y$ direction within one period time. 

\subsection{RLBL invariant calculation}
The Rudner-Lindner-Berg-Levin (RLBL) invariant plays a central role in Floquet phases that cannot be solely explained by the Chern number. It is initially introduced for explicit time-dependent Hamiltonian. For discrete-time quantum walks, equivalent time-dependent Hamiltonians need to be constructed. 

The RLBL invariant $W$ is defined as
\begin{equation}\label{eq:2D_RLBLWinding}
W[U_{\varepsilon}] = \frac{1}{8\pi^2}\int dtdk_xdk_y\mathrm{Tr}(U_{\varepsilon}^{-1} \partial_t U_{\varepsilon}[U_{\varepsilon}^{-1} \partial_{k_x} U_{\varepsilon},U_{\varepsilon}^{-1} \partial_{k_y} U_{\varepsilon}]).
\end{equation}
Here $U_{\varepsilon}$ is given by
\begin{equation}
  U_{\varepsilon} =
    \begin{cases}
      U(k_x,k_y,2t) & \text{if $0\leq t\leq T/2$}\\
      V_{\varepsilon}(k_x,k_y,2T-2t) & \text{if $T/2\leq t\leq T$}
    \end{cases}       
\end{equation}
where $V_{\varepsilon}(k_x,k_y,t) = \e^{-\iu H_{\mathrm{eff}}t}$, $H_{\mathrm{eff}} = \frac{\iu}{T}\mathrm{log}U(k_x,k_y,T)$, $\varepsilon$ is the quasienergy band gap, and we take $T = 1$. 

To construct a time-dependent Hamiltonian, we consider a square lattice of a unit cell, each containing four sites, denoted by $\{a\!\uparrow, a\!\downarrow, b\!\uparrow, b\!\downarrow \}$ where $\{a,b\}$ are the coin states and $\{\uparrow,\downarrow\}$ are the pseudospin state as shown in the main text. These sites are identified with the states of the walker as 
\begin{align}
c^{\dagger}_{x,y,a_s } = \ket{x,y,a_s}, \quad c^{\dagger}_{x,y,b_s } = -\iu \ket{x,y,b_s},
\end{align}
where $s = \{\uparrow, \downarrow\}$ and $x,y$ are 2D lattice sites. To simplify the notion, we define $\mathbf{\hat{c}}_{x,y,a} = \begin{pmatrix}\hat{c}_{x,y,a\!\uparrow}\\\hat{c}_{x,y,a\!\downarrow}\end{pmatrix}$ and $\mathbf{\hat{c}}^{\dagger}_{x,y,a} = \left(\hat{c}^{\dagger}_{x,y,a\!\uparrow},\hat{c}^{\dagger}_{x,y,a\!\downarrow} \right)$ and similarly for $\mathbf{\hat{c}}_{x,y,b}$ and $\mathbf{\hat{c}}^{\dagger}_{x,y,b}$. 
Based on the method proposed in Ref.\cite{asboth2015edge}, we map our 2D topological quantum walk [Eq.~\eqref{eqS:2D_topolo}] into a nearest-neighbor hopping Bloch Hamiltonian on the lattice
\begin{align}\label{eq:2DtimeHamiltonian}
\begin{split}
H(t) =   u(t) (\tau_x\otimes\sigma_0) + U_y(t,\phi)(\tau_z\otimes\sigmay) + U_x(t,\theta)(\tau_z\otimes\sigmaz) + S_x(t) \begin{pmatrix}
0 & \e^{\iu k_x} \\
\e^{-\iu k_x} & 0
\end{pmatrix} \otimes \sigma_0 + S_y(t) \begin{pmatrix}
0 & \e^{\iu k_y} \\
\e^{-\iu k_y} & 0
\end{pmatrix} \otimes \sigma_0,
\end{split}
\end{align}
where $u(t)$ corresponds to the coin space rotation, $U_{x(y)}$ correspond to the gauge fields in $x(y)$-direction, and $S_{x(y)}(t)$ combined with $u(t)$ to realize hopping in $x(y)$ direction. 

\begin{figure}[h] 
	\centering
	\includegraphics[width=1\linewidth]{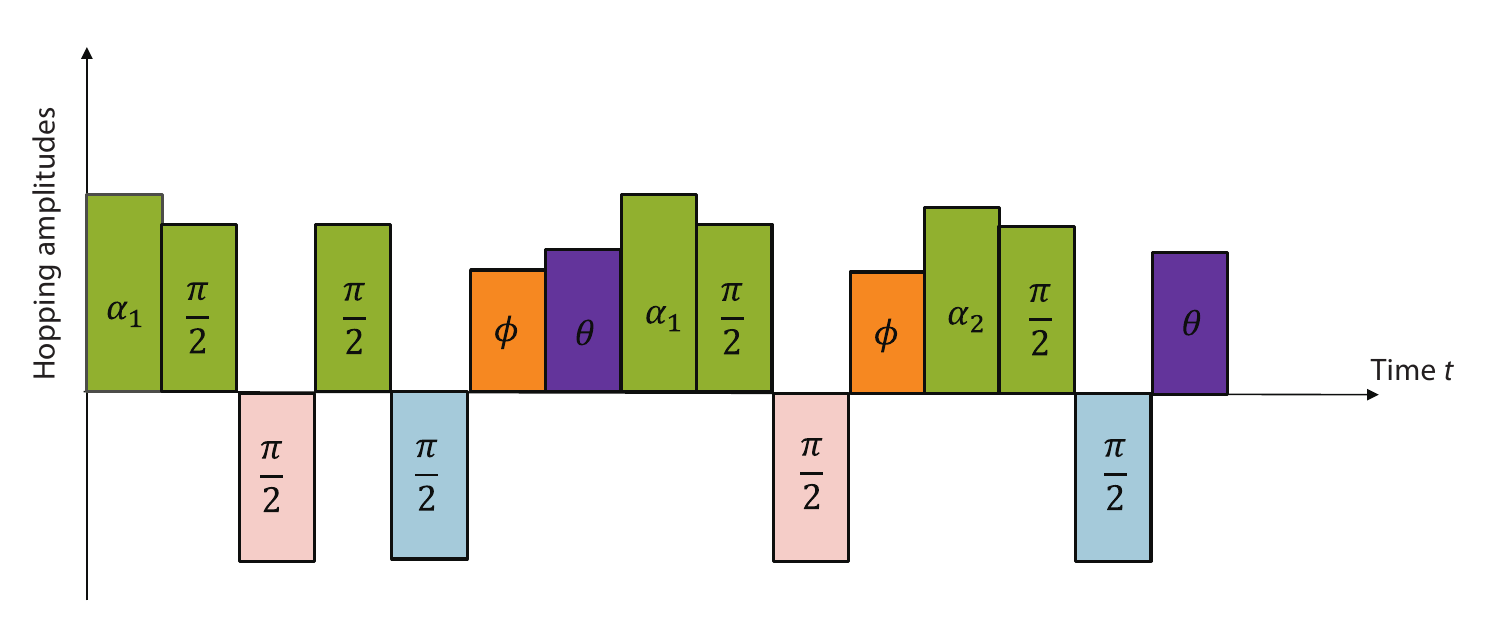}
	 \caption{%
	 	\textbf{Drive sequence for the lattice Hamiltonian.}
	 	The time-dependent Hamiltonian is realized by a sequence of 15 non-overlapping pulses. Green color represents $u(t)$, pink color represents $S_y(t)$, blue color represents $S_x(t)$, purple and orange represent gauge fields in $x$ and $y$ direction.
	 	}\label{figs:TimeH}
\end{figure}

To realize the 2D quantum walk [Eq.~\eqref{eqS:2D_topolo}], we can use a non-overlapping sequence of pulses where at any time, only one type of pulse is switched on shown in Fig.~\ref{figs:TimeH}. A pulse of $u(t)$ of area $\pi/2$ followed by a pulse of $S_x$ of area $-\pi/2$ realize the hopping in $x$-direction. Similarly, we can realize hopping in $y$-direction. A pulse of $u(t)$ of area $\alpha_{1(2)}$ can realize the rotation of angle $\alpha_{1(2)}$ in coin space. A pulse of $U_{x(y)}$ of area $\theta (\phi)$ can realize rotation angle of $\theta (\phi)$ by the synthetic gauge fields.

The pulse sequence realizing Eq.~\eqref{eqS:2D_topolo} consists of 15 pulses. We assume the period of time is unity, and each pulse has a width of $1/15$. Each pulse is explicitly defined as
\begin{align}\label{eq:G_time}
\begin{split}
G(t,i) = 15\times \Bigl\{ (\mathrm{sgn}[t-i]+1)/2 - \mathrm{sgn}[t-(i+1/15)]+1)/2 \Bigr\}.
\end{split}
\end{align}
Here $\mathrm{sgn}[x]$ is a sign function and $i$ ranges from 0 to $14/15$ with step size $1/15$. 
We can define a discrete set $G(t) \equiv \left\{G(t,i) \mid i\right\}$ and $G[m]$ represents the $m$-th element in the set $G(t)$.
The hopping amplitudes are explicitly given by
\begin{align}\label{eq:HoppingAmplitude}
\begin{split}
u(t) &= \alpha_1G[1] + \frac{\pi}{2}G[2] + \frac{\pi}{2}G[4] + \alpha_1G[8] + \frac{\pi}{2}G[9] + \alpha_2G[12] + \frac{\pi}{2}G[13],\\
U_x(t) &= \theta (G[7] + G[15]),\\
U_y(t) &= \phi (G[6] + G[11]),\\
S_x(t) &= -\frac{\pi}{2}(G[5] + G[14]),\\
S_y(t) &= -\frac{\pi}{2}(G[3] + G[10]).
\end{split}
\end{align}

From this time-dependent Hamiltonian, we are able to calculate the RLBL invariant. 

\section{Spectral flow of quantum walk}\label{S5}

We present a complementary method for calculating the RLRL invariant of our quantum walk using the spectral flow method~\cite{asboth2017spectral}.

The spectral flow method relates the RLBL invariant of a certain quasienergy gap to the spectral flow of the same gap induced by tuning the magnetic flux from one commensurate value to the next.
The method was originally applied to magnetic quantum walks~\cite{sajid2019creating,asboth2017spectral,lin2023manipulating} featuring inhomogeneous U(1) gauge fields. To apply the method here, we introduced a small fictitious magnetic flux in our system. To obtain the correct RLBL invariant, we select a magnetic flux that is small enough. Now, the modified pseudo-magnetic 2D quantum walk is given by
\begin{align}
U_{\mathrm{2D}}^{\mathrm{modified}}(k_x,k_y) = S_x(k_x)C_2F(B/2)S_y(k_y)C_1S_x(k_x)F(B/2)S_y(k_y)C_1.
    \label{eq:2D_magnetic}
\end{align}
Here, $F(B)$ is the magnetic-field operator read as $F(B) = \mathrm{exp}[\iu \sigmaz (\beta\lfloor\hat{x}/q\rfloor+ B \hat{x})]$, with $\hat{x}$ being the lattice position operator along the $x$ axis. $\beta = 2\pi/s$ being the additional fictitious magnetic field and $\lfloor\hat{x}/q\rfloor$ is the greatest integer less than or equal to $x/q$ indexing the magnetic unit cells.  
$B = 2\pi p/q$ is the flux per plaquette, where $p$ and $q$ are coprime integers. In our case, we choose $p/q = 1/100$ and $s = 15$, and $B/2$ in the Eq.~\eqref{eq:2D_magnetic} accounts for the fact that the quantum walker can move two unit cells in one period. 
Following the spectral flow method, the RLBL invariant  $W_{\tilde{E}}$ of the $\tilde{E}$ gap is given by
\begin{align}
 W_{\tilde{E}} = \frac{1}{2\pi} \biggl\{\sum_{j}^{2sq}E_j(1/s,B,\tilde{E}) - \sum_{j}^{2sq}E_j(0,B,\tilde{E}) \biggr\},
    \label{eq:2D_magneticRLBL}
\end{align}

Where $U' = \e^{\iu \tilde{E}}U_{\mathrm{2D}}^{\mathrm{modified}}$, $\e^{\iu \tilde{E}}$ shifts the quasienergy spectrum by $-\tilde{E}$.




\bibliographystyle{apsrev4-1-fixspace} 

%